# Temporal Dynamics and Security Assessment of SRT Protocol: A 96-Day Longitudinal Analysis of Production Streaming Infrastructure

Ahmed F. K. Koysha * and Ferdi Sönmez

Department of Computer Engineering, İstanbul Aydın University, Istanbul, Turkey
Email: afouadkoysha@stu.aydin.edu.tr (A.F.K.K); ferdisonmez@aydin.edu.tr (F.S.)
*Corresponding author

***Abstract*—This study presents a 96-day longitudinal analysis of Secure Reliable Transport (SRT) protocol behavior in a production streaming environment between Iraq and Turkey, examining 262,672 transport metric samples and 5,033,104 system resource measurements collected between October 2025 and January 2026. The research addresses temporal patterns, system-resource relationships, attack vulnerability, and anomaly detection. Results reveal significant temporal variation: Negative Acknowledgment (NAK_rate) exhibits 7.6-fold hourly variation and 11.5-fold daily variation ($p < 0.001$), though with small effect sizes ($\eta^2 < 0.04$), indicating that temporal period explains less than 4% of total variance. Central Processing Unit (CPU) utilization emerges as the strongest system-level correlate of transport variation ($r = 0.617$ with NAK_rate). Controlled User Datagram Protocol (UDP) flood experiments at 500 packets per second cause CPU increases of 176%–806%, while Secure Reliable Transport (SRT) transport metrics change by only ±13%, demonstrating protocol resilience through the Automatic Repeat Request (ARQ) mechanism. An unexpected anomaly reaching 97.7% CPU enabled signature-based differentiation between attack types. A *Z*-score anomaly detection method ($3\sigma$) outperforms fixed thresholds, detecting all 7 security events. This study contributes the longest temporal characterization of SRT metrics reported to date, the first system-SRT correlation analysis, and the first controlled security assessment of SRT infrastructure. These findings provide actionable guidance for capacity planning, security hardening, and multi-metric monitoring of production SRT deployments.**

## I. INTRODUCTION

The exponential growth of live streaming services has placed unprecedented demands on transport protocols to maintain quality of service across diverse and unpredictable network conditions. The global live streaming market, valued at over $70 billion in 2024, relies increasingly on protocols capable of delivering low-latency, high-reliability transmission over public internet infrastructure [1]. As audiences now expect near-instant interaction during live sports, auctions, and interactive broadcasts, the limitations of legacy protocols such as Real-Time Messaging Protocol (RTMP) have become increasingly apparent, driving adoption of transport-layer solutions designed for unreliable network paths. Among these solutions, the Secure Reliable Transport (SRT) protocol has established itself as a leading technology for professional broadcasting since its open-source release by Haivision [2]. By combining Advanced Encryption Standard (AES) encryption with Automatic Repeat Request (ARQ) error correction, SRT enables broadcast-quality delivery even over connections affected by significant packet loss and latency variability. The protocol has gained widespread adoption across broadcast, enterprise, and consumer streaming applications, yet fundamental questions about its operational behavior in production environments remain open.

Despite growing academic interest in SRT, the existing literature exhibits several critical gaps that limit practical understanding of protocol behavior. First, the temporal dimension of SRT performance remains largely unexplored; while time series methods have been applied to related domains such as Voice over Internet Protocol (VoIP) quality forecasting [3] and general streaming traffic prediction [4], no study has characterized the hourly, daily, or weekly patterns in SRT transport metrics. Second, prior SRT studies have been conducted predominantly in laboratory testbed environments that cannot capture the traffic variability, network dynamics, and operational conditions present in production deployments [5, 6]. Third, the relationship between system resource utilization and SRT transport performance has not been systematically investigated; existing resource-quality studies focus on Quality of Experience metrics rather than transport-layer indicators [7]. Fourth, while Distributed Denial-of-Service (DDoS) attack research has matured substantially [8], SRT-specific security assessment remains absent from the literature,

Manuscript received January 19, 2026; revised March 5, 2026; accepted March 19, 2026; published August 24, 2026.

leaving operators without guidance on vulnerability thresholds or attack signatures specific to SRT infrastructure.

A particularly consequential gap is the limited duration of existing studies. The longest reported SRT performance study spans approximately one week [6], while most experiments measure protocol behavior over hours or days. This temporal constraint precludes detection of weekly cyclical patterns, long-term trend variations, and seasonal effects that may substantially influence operational planning. Reviews of Quality of Experience studies have repeatedly identified the need for longitudinal approaches capturing behavior across extended periods [9], yet SRT research has not responded to this methodological imperative. Comprehensive workload analyses of live streaming systems have demonstrated significant temporal structure at multiple scales, suggesting that similar patterns likely exist in SRT transport metrics awaiting systematic characterization [10].

To address these gaps, this study poses 2 research questions.

**RQ1:** What temporal patterns exist in SRT protocol transport metrics at hourly, daily, and weekly granularities in a production streaming environment?

**RQ2:** What is the relationship between system resource utilization Central Processing Unit (CPU), Random Access Memory (RAM), network) and SRT transport performance metrics?

**H1:** Posits that SRT metrics exhibit significant temporal patterns.

**H2:** Posits that system resources correlate with transport performance.

These two questions collectively address a core methodological challenge in SRT research: bridging the gap between short-duration laboratory evaluations and production-scale operational understanding. Existing studies provide valuable protocol-level insights but lack the temporal depth, system-integration perspective needed to guide real-world deployment decisions.

This study presents a 96-day longitudinal analysis of SRT protocol behavior conducted between October 14, 2025 and January 18, 2026 in a production streaming environment. The research collected 262,672 SRT transport metric samples at approximately 30 s intervals and 5,033,104 system resource measurements at 1 s resolution, representing the most extensive empirical characterization of SRT behavior reported in the literature. The methodology combines passive observational monitoring for temporal pattern analysis with controlled experimental interventions for security assessment, enabling both ecological validity through production data and experimental rigor through isolated attack testing. Statistical analysis employs Analysis of Variance (ANOVA), Kruskal-Wallis, and Mann-Whitney tests for temporal pattern significance; Pearson correlation for metric relationships; and additive time series decomposition with Autocorrelation Function (ACF) and Partial Autocorrelation Function (PACF) analysis for cyclical structure identification.

The study offers 4 contributions. First, to our knowledge, it represents the longest reported monitoring of SRT transport behavior—spanning 96 days compared to roughly one week in prior work—and uncovers weekly cyclical patterns suitable for Seasonal Autoregressive Integrated Moving Average (SARIMA)-based forecasting. Second, it provides the first systematic examination of how system resources relate to SRT performance, with CPU utilization emerging as the strongest correlate of transport variation ($r$ = 0.617). Third, it introduces a controlled security evaluation of SRT infrastructure, showing that attack traffic well below conventional DDoS thresholds canproduce measurable impact. Finally, by grounding the analysis in a production deployment rather than a laboratory testbed, it captures operational dynamics that controlled environments typically miss.

The paper proceeds as follows: Section II surveys the relevant literature; Section III details the data collection, statistical methods, and attack protocols; Section IV reports the results and discusses their implications and limitations; and Section V presents the conclusions.

## II. Literature Review

The proliferation of live streaming services has catalyzed extensive research into transport protocol optimization, Quality of Service (QoS) monitoring, and security assessment of streaming infrastructure. The SRT protocol, released as open-source by Haivision in 2017, has emerged as a critical technology for professional broadcasting and content delivery networks. This literature review examines existing research across 6 interconnected domains: SRT protocol performance studies, temporal analysis methodologies for streaming quality, system resource impact on transport metrics, streaming infrastructure security, identification of research gaps that motivate the present study, and recent advances in real-time transport protocols.

### *A. SRT Protocol Performance Studies*

Research on SRT protocol performance has expanded significantly since its open-source release. Several studies target transport-layer optimization: machine learning-enhanced codec selection using K-means clustering and K-nearest neighbors achieved 98% prediction accuracy in reducing latency and jitter [11], while adaptive rate control based on periodic network feedback improved latency under varying network conditions [6]. Others extend SRT to specific deployment contexts, including cellular handover [12], the Network Video Interface (NVI) protocol for professional live production [13], and Low-Latency Dynamic Adaptive Streaming over HTTP (LL-DASH) systems achieving end-to-end delay under 2 s [5]. Hybrid designs have also been explored, notably SRT over QUIC datagrams leveraging QUIC's TLS-based security and congestion control [14]. Empirical evaluation of such hybrids remains limited: QUIC suffers throughput reductions of up to 45% on high-speed networks [16], its SRT specifications remain at the Internet-Draft stage, and standalone SRT continues to dominate production deployments.

Despite these advances, a critical limitation pervades existing SRT research: experimental durations typically span hours to days. The longest documented SRT performance study prior to the present work spans approximately one week [6], preventing identification of weekly patterns, seasonal trends, or long-term performance drift. Our 96-day study addresses this gap, providing the most extended continuous SRT monitoring reported in the literature.

### *B. Temporal Analysis in Streaming Quality*

Time series analysis has been extensively applied to streaming quality prediction and network traffic characterization. Deep learning dominates recent work: Bidirectional Long Short-Term Memory (LSTM) and Extra Tree models outperform traditional techniques for viewer count prediction [1], while Vector Autoregression (VAR) models and recurrent architectures have been used for VoIP quality forecasting in mobile networks [3, 15], capturing relationships among Quality of Service and Quality of Experience (QoS/QoE) metrics. Classical statistical models remain competitive: comparison of Autoregressive Integrated Moving Average (ARIMA) and Fractional ARIMA (FARIMA) models showed that streaming traffic exhibits long-range dependencies amenable to time series modeling [4], later extended through hybrid FARIMA/Generalized Autoregressive Conditional Heteroskedasticity (GARCH) models for dynamic bandwidth allocation [17].

Decomposition-based methods have proven equally effective, including seasonal decomposition with Dynamic Time Warping for pattern recognition [18], Quartile-Based Seasonality Decomposition for large-scale telecom KPI datasets [19], packet-flow feature extraction for QoE estimation [20], and LSTM modeling of nonlinear QoE dynamics [21].

These studies, however, predominantly address application-layer QoE rather than transport-layer protocol behavior. The systematic review by Cieplinska *et al.* [9] found that long-term QoE assessment studies remain rare, most spanning weeks rather than months, and none specifically targets SRT transport metrics. Our study applies time series decomposition with period = 7 to SRT metrics, revealing weekly seasonality ($p < 0.001$) and establishing ACF/PACF patterns suitable for SARIMA modeling.

### *C. System Resource Impact on Streaming Quality*

Research correlating system resource utilization with streaming quality has established relationships enabling predictive modeling and adaptive strategies. Statistical analysis and machine learning regression applied to Web Real-Time Communication (WebRTC) sessions showed strong correlation between application-layer parameters and Mean Opinion Score (MOS) [7], and AI-based link quality prediction has been used to maximize QoE in mobile video streaming [22]. Related work has modeled RTT time series for anomaly detection [23], examined video quality under time-varying loads, finding weak correlations that require load-adjusted models [24], and integrated digital twin analytics for dynamic resource allocation [25].

These analyses nonetheless rely on limited datasets or short observation windows [22], and none has examined the relationship between system resource utilization and SRT transport metrics. Our study addresses this gap through 5,033,104 system resource measurements correlated with 262,672 SRT samples, revealing that CPU usage correlates strongly with NAK_rate ($r = 0.617$) and network throughput ($r = 0.677$), while RAM correlates moderately with RTT_avg ($r = 0.509$).

### *D. Streaming Infrastructure Security*

DDoS attacks on streaming infrastructure present significant challenges for service availability. A comprehensive survey of defense mechanisms established taxonomies for UDP flood, SYN flood, and application-layer attacks [8]; UDP floods target the transport layer by overwhelming server resources, causing degradation or denial of service. Detection research has increasingly turned to machine learning, including unsupervised inference of QoE degradation among residential users [26] and change point detection with Generalized Autoregressive Conditional Heteroskedasticity (GARCH) modeling to identify traffic volatility indicative of anomalous behavior [27]. Satellite streaming raises further concerns: Starlink analysis revealed that abrupt variability and frequent handovers create vulnerabilities current adaptive mechanisms inadequately address [28], and Geostationary Earth Orbit (GEO) latency degrades QoE through disrupted scheduling [29].

SRT-specific vulnerability assessment nonetheless remains underexplored: no study has quantified the impact of controlled attack scenarios on SRT infrastructure. Our 6 controlled UDP flood experiments demonstrate that even low-intensity attacks (500 pps) cause CPU increases up to 806% above baseline, while medium-intensity attacks reach 1348%; an unexpected network anomaly reaching 97.7% CPU with a 449% traffic increase further demonstrates real-world exposure. This focus follows growing attention to low-rate DDoS attacks as a distinct threat category, operating at hundreds to thousands of packets per second rather than volumetric floods exceeding 100 Gbps and exploiting protocol-specific weaknesses while evading conventional detection [30], and to machine learning methods reporting above 95% detection accuracy from flow-level and statistical anomaly features [31].

### *E. Research Gap Summary*

The synthesis of existing literature reveals critical gaps in understanding SRT protocol behavior under production conditions. Table I and Table II contextualize this study relative to prior research, highlighting our extended 96-day monitoring, collection of 262,672 SRT transport samples, and the introduction of a controlled attack assessment on SRT infrastructure.

TABLE I. COMPARATIVE SUMMARY OF SRT PROTOCOL STUDIES

| Study | Duration | Dataset Size | Metrics Scope | Security Eval. | Environment |
|---|---|---|---|---|---|
| [6] | ~1 week | Not reported | Bitrate, latency, packet loss | None | Laboratory testbed |
| [5] | Hours | ~500 samples | Throughput, latency, DASH segments | None | Laboratory testbed |
| [16] | Hours | Not reported | QUIC throughput, RTT | None | Controlled network |
| [32] | Hours | Not reported | P2P relay, latency, throughput | None | IPv6 testbed |
| **This study (2026)** | **96 days** | **262,672 + 5,033,104** | **5 SRT + 4 system metrics** | **6 controlled attacks + anomaly** | **Production (Iraq–Turkey)** |

TABLE II. SUMMARY OF LITERATURE GAPS AND STUDY CONTRIBUTIONS

| Gap Area | Literature Limitation | Key Studies | Our Contribution |
|---|---|---|---|
| Study Duration | Most SRT studies span hours to days; longest prior study ~1 week | [6, 11] | 96-day continuous monitoring (262,672 samples) |
| Temporal Analysis of SRT | Time series methods applied to QoE, not SRT transport metrics | [3, 4] | Weekly seasonality, ACF/PACF, decomposition for SRT |
| Hourly/Daily Patterns | No characterization of SRT diurnal or weekly cycles | [1, 17] | 7.6× NAK variation hourly; 5.1× weekday-weekend differential |
| System-SRT Correlation | Resource-quality studies focus on QoE, not transport layer | [7, 23] | CPU-NAK ($r = 0.617$); CPU-throughput ($r = 0.677$) correlation |
| SRT Security Assessment | DDoS research lacks SRT-specific vulnerability quantification | [8, 26] | Controlled attacks: +806% CPU at 500 pps; anomaly: 97.7% CPU |
| Production Environment | Laboratory testbeds dominate; limited real-world validation | [5, 13] | Continuous production deployment with real viewer traffic |
| Protocol Comparison | No longitudinal comparison of SRT with emerging alternatives | [14, 29] | Contextualizes SRT within QUIC/MoQ landscape |

The identified gaps reflect a fundamental tension in streaming research between controlled laboratory evaluation and production-scale validation. While laboratory studies [5, 11] provide precise measurement conditions, they cannot capture the complexity of temporal patterns, resource interactions, and security threats that emerge in operational environments. The systematic review in Ref. [9] explicitly calls for extended-duration studies that integrate objective metrics with real-world deployment conditions.

### *F. Recent Advances in Real-Time Transport Protocols (2024–2025)*

Since this study's design, the real-time transport landscape has rapidly evolved. While the IETF Media over QUIC (MoQ) group advances QUIC-based media delivery [33] and hybrid SRT-over-QUIC models have been proposed, QUIC implementations remain in early stages with higher protocol overhead than native UDP solutions [14]. Consequently, standalone SRT adoption continues to surge; it is now utilized by 68% of surveyed broadcasters compared to 56% for RTMP [34], driving new architectural proposals like IPv6 decentralized chain-relays [32]. Despite this widespread deployment, no longitudinal production monitoring study has appeared in the literature. Furthermore, as DDoS attack frequencies increased by 30% in Ref. [35] and edge environments face growing vulnerabilities to low-rate protocol-level exploits [36], the need to evaluate the security and stability of distributed SRT deployments under production conditions has become increasingly urgent.

While SRT-QUIC integration represents a promising future direction, the present study focuses on standalone SRT because: (a) it remains the dominant standard for live broadcasting; (b) SRT-over-QUIC is still at the Internet-Draft stage with limited deployment; and (c) understanding baseline SRT behavior provides essential groundwork for evaluating future hybrid architectures.

## III. MATERIALS AND METHODS

### *A. Research Design and Data Collection Infrastructure*

This study employs a mixed-methods research design that integrates longitudinal observational analysis with controlled experimental validation to comprehensively characterize the temporal dynamics and security vulnerabilities of the Secure Reliable Transport (SRT) protocol. The research framework was specifically designed to address 5 critical gaps identified in the existing literature: the absence of long-term empirical studies spanning multiple months, the lack of systematic temporal pattern analysis for SRT transport metrics, the unexplored relationship between system resources and protocol performance, the insufficient security assessment of SRT under attack conditions, and the predominance of laboratory-based evaluations that fail to capture production environment dynamics.

The 96-day observation period was strategically selected to capture complete seasonal cycles, multiple weekly patterns, and sufficient temporal variation for robust statistical analysis. This duration substantially exceeds the typical hours-to-days span of existing SRT studies, enabling the identification of long-term trends and cyclical behaviors that shorter studies cannot detect. The research methodology combines 2 complementary components: passive continuous monitoring of a production streaming infrastructure for temporal characterization, and active controlled attack experiments in an isolated laboratory environment for security assessment. This dual approach ensures both ecological validity through real-world data collection and experimental rigor through controlled intervention testing.

The data collection infrastructure was deployed on a dedicated streaming server operating within a production broadcast environment. The hardware and software configuration of the monitoring server is detailed in Table III.

TABLE III. SERVER HARDWARE AND SOFTWARE CONFIGURATION

| Component | Specification |
|---|---|
| Device | I-Life ZED Air CX3 (laptop) |
| CPU | Intel Core i3-5005U @ 2.00 GHz, 2 cores / 4 threads |
| RAM | 3.2 GB DDR3 |
| OS | Ubuntu 25.10 (kernel 6.17.0-8-generic, x86_64) |
| SRT Software | srt-live-transmit v1.5.4 (libSRT 1.5.4, OpenSSL) |
| Network | 1 Gbps Ethernet (192.168.1.0/24) |
| Storage | 111.8 GB SSD |
| Monitoring DB | InfluxDB 2.x with Grafana 10.x |

The server was configured to continuously capture SRT transport metrics and system resource utilization throughout the study period. An InfluxDB time-series database (version 2.x) served as the primary storage system, selected for its optimized handling of high-frequency temporal data and efficient compression algorithms suitable for long-term metric storage.

SRT metrics were sampled at a mean interval of 31.6 s (SD = 2.3 s), with minor variation due to system scheduling overhead, yielding 262,672 data points over the 96-day collection period spanning from October 14, 2025 to January 18, 2026. System resource metrics were collected at 1 s intervals, producing 5,033,104 samples over the 58.25-day overlapping period from November 21, 2025 (10: 02: 48) to January 18, 2026 (23: 59: 59). An overview of both datasets is provided in Table IV.

TABLE IV. DATASET OVERVIEW AND COLLECTION PARAMETERS

| Dataset | Samples | Start | End | Days |
|---|---|---|---|---|
| SRT Metrics | 262,672 | 10/14/2025 11:54 AM | 1/18/2026 11:59 PM | 96 |
| System Stats | 5,033,104 | 11/21/2025 10:02 AM | 1/18/2026 11:59 PM | 58 |

The system resource dataset falls entirely within the SRT monitoring period, providing complete temporal overlap for all 58 days of system data. Cross-correlation analysis utilized this full overlap period, ensuring that all system-SRT relationships are computed from concurrent measurements.

The differing sampling rates for SRT transport metrics (30 s intervals) and system resources (1 s intervals) reflect the distinct temporal characteristics of each metric category. SRT statistics are computed as cumulative values over measurement windows; shorter intervals would yield statistically unstable estimates due to insufficient packet counts within each window. System resource metrics, conversely, exhibit rapid fluctuations requiring higher temporal resolution to capture transient events such as attack-induced CPU spikes. Data was exported from InfluxDB in CSV format using Flux queries, with timestamps preserved in ISO 8601 format with UTC timezone standardization.

### B. SRT Transport and System Resource Metrics

Five primary SRT transport metrics were continuously monitored throughout the study period, each providing distinct insights into stream quality and transmission reliability.

The Negative Acknowledgment Rate (NAK_rate) quantifies the proportion of packets requiring retransmission requests, serving as a direct indicator of packet loss severity. Values approaching zero indicate optimal transmission, while elevated rates signal network congestion or quality degradation.

Round-Trip Time Average (RTT_avg) measures the mean bidirectional latency between sender and receiver, with lower values indicating more responsive connections. The Round-Trip Time Standard Deviation (RTT_std) captures network jitter by measuring RTT variability. Buffer variance quantifies fluctuations in the receiver buffer level, with higher values indicating more variable playout buffering that may compromise playback smoothness. The cumulative dropped packet counter (dropped_total) tracks irrecoverable packet losses.

System resource monitoring encompassed 4 metrics capturing server computational and network characteristics. CPU utilization (cpu_usage) was sampled at 1 s intervals. Network throughput was captured bidirectionally: received bandwidth (mbps_recv) reflecting the incoming stream bitrate, while transmitted bandwidth (mbps_sent) representing control traffic and acknowledgments.

Temporal synchronization between SRT and system datasets presented a methodological challenge due to their differing sampling rates. The adopted approach aggregated both datasets to hourly intervals using mean values for correlation analysis, preserving sufficient temporal resolution for pattern detection while ensuring measurement alignment. This aggregation strategy also reduced noise in high-frequency measurements and enabled statistically robust correlation estimation across the 58-day overlap period.

### C. Statistical Analysis and Time Series Methods

Comprehensive temporal features were derived from the raw timestamp data to enable multi-scale pattern analysis. The hour of day (0−23) was extracted to characterize diurnal patterns in network behavior. Day of week (0−6, where 0 represents Monday) and the corresponding day name facilitated weekly cycle analysis and weekend/weekday comparisons. Week number (1−52) and month (1−12) supported longer-term trend identification across the study period. A binary weekend indicator (0 for weekdays, 1 for weekends) enabled direct statistical comparison between workweek and weekend performance. Calendar date extraction supported daily aggregation for time series decomposition. These temporal features were systematically applied to segment the data for hourly pattern analysis, daily distribution characterization, and weekly trend examination.

The statistical analysis framework employed a combination of parametric and non-parametric tests, selected based on the distributional characteristics of each metric. A stringent significance threshold of $\alpha = 0.001$ was adopted throughout to minimize Type I error probability given the large sample sizes.

Preliminary normality assessment using Shapiro-Wilk tests confirmed non-normal distributions for all SRT and system metrics ($p < 0.001$ in all cases), with pronounced right-skewed distributions characteristic of network performance data. This motivated the use of both parametric tests (ANOVA, for comparison with prior literature) and non-parametric alternatives (Kruskal-Wallis, Mann-Whitney U) as the primary analytical framework, with concordance between methods reported throughout.

One-way Analysis of Variance (ANOVA) was applied to test for significant differences in metric means across the 24 h of the day. The null hypothesis H0: $\mu_1 = \mu_2 = \ldots = \mu_{24}$ was tested for NAK_rate, RTT_avg, and buffer_variance. Given the non-normal distributions exhibited by most metrics, the Kruskal-Wallis H-test was employed to compare medians across the seven days of the week. This non-parametric alternative to one-way ANOVA tests the null hypothesis that samples from different groups originate from the same distribution.

The Mann-Whitney U-test facilitated pairwise comparison between weekend and weekday performance, testing whether the 2 groups differ in their central tendency.

Classical additive time series decomposition was applied to isolate the constituent components of SRT metric temporal variation. The additive model $Y(t) = T(t) + S(t) + R(t)$ decomposes the observed time series $Y(t)$ into trend $T(t)$, seasonal $S(t)$, and residual $R(t)$ components. A 7-day period was selected for seasonal decomposition based on the strong weekly cyclical patterns observed in preliminary analysis and the practical relevance of weekly business cycles to streaming traffic patterns.

The decomposition was implemented using the stats models Python library seasonal decompose function with the additive model specification.

The Autocorrelation Function (ACF) was computed to detect temporal dependencies in daily aggregated NAK_rate values across 30 lag periods. The ACF measures the correlation between observations separated by $k$ time units. The Partial Autocorrelation Function (PACF) isolated direct temporal effects by controlling for intermediate lags.

Pearson product-moment correlation coefficients were computed to quantify linear relationships between SRT metrics.

Cross-correlation analysis between system resources and SRT metrics was conducted using hourly aggregated data over the 58-day overlap period.

This aggregation strategy addressed the sampling rate mismatch between 1 s system data and 30 s SRT data while providing sufficient temporal resolution for meaningful correlation estimation.

### *D. Attack and Anomaly Detection Protocols*

Controlled attack experiments were conducted in an isolated laboratory network environment to assess SRT protocol vulnerability without impacting production systems or external networks. The laboratory was configured as a private network segment (192.168.1.0/24) physically disconnected from the internet, ensuring that attack traffic remained contained. The attack workstation (192.168.1.83) targeted the SRT server on port 4000. All experiments were conducted on December 28, 2025, with comprehensive system monitoring capturing attack impacts at 1 s resolution.

Attack traffic was generated using the Python Scapy library, which enables low-level packet crafting with precise control over protocol headers. 2 attack types were evaluated: SRT-specific UDP floods with crafted SRT control headers, and generic UDP floods without protocol-specific payloads. 6 attack sessions were executed between 20:10 and 21:07 local Turkey time (UTC+3), corresponding to 17:10–18:07 UTC: three low-intensity SRT floods at 500 packets per second (attacks 1–3, 5 min each), and three medium-intensity UDP floods at approximately 1000+ packets per second (attacks 4–6, 10 min each).

Each attack session was conducted for a fixed duration: Attacks 1–3 (SRT Low, 500 pps) lasted approximately 5 min each, while Attacks 4–6 (UDP Medium, 1000 pps) lasted 8–10 ms each, with 2–3 ms intervals between sessions for system recovery monitoring.

A 30 min baseline period prior to attack initiation established reference performance levels with mean CPU utilization of 1.81%, RAM at 25.24%, and network throughput of 8.08 Mbps. Attack impact was quantified using the percentage change formula: Impact% = (V_attack−V_baseline)/V_baseline) × 100, where V_attack represents the maximum metric value during the attack period and V_baseline represents the pre-attack mean.

An unexpected network anomaly occurred on December 29, 2025 at 07:07:56 UTC (10:07:56 local Turkey time), providing an opportunity to validate anomaly detection methodology against a naturally occurring event.

The detection approach employed a threshold-based criterion: CPU utilization exceeding 10% (approximately 10 standard deviations above the mean of 1.52%) was flagged as anomalous. Software and Reproducibility.

The data preprocessing pipeline addressed several challenges inherent to InfluxDB Comma-Separated Values (CSV) exports. Raw exports contained 3 header rows of metadata which were skipped during import. Timestamp validation filtered rows where the _time field contained the International Organization for Standardization (ISO) 8601 'T' separator, excluding malformed entries. Timestamps were parsed using mixed format detection with Coordinated Universal Time (UTC) timezone normalization, then converted to timezone-naive format for analysis. The original long-format data structure, with _field and _value columns representing variable names and their corresponding measurements, was pivoted to wide format using pandas pivot_table with timestamp as index and field names as columns. Numeric conversion was applied to all metric columns with coercion of non-numeric values to NaN, followed by removal of rows with missing values in critical columns. The complete preprocessing pipeline ensured data integrity while preparing a structured dataset suitable for statistical analysis.

All analyses were implemented in Python 3.10+ using the following core libraries: pandas 2.0+ for data manipulation and preprocessing, numpy 1.24+ for numerical computation, scipy 1.11+ for statistical hypothesis testing, statsmodels 0.14+ for time series analysis and decomposition, matplotlib 3.7+ for visualization, and seaborn 0.12+ for statistical graphics. InfluxDB 2.x served as the time-series database platform, with Flux query language used for data extraction. Google Colab provided the computational environment for analysis execution. To support reproducibility, the complete analysis code, preprocessing scripts, and figure generation routines are available upon reasonable request from the corresponding author. The anonymized dataset and analysis notebooks will be deposited in a public repository upon manuscript acceptance.

## IV. Results and Discussion

### A. Dataset Overview and Descriptive Statistics

The longitudinal monitoring campaign yielded 2 comprehensive datasets spanning the 96-day study period, as presented in Table IV. The SRT transport metrics dataset comprised 262,672 samples collected continuously from October 14, 2025 (11:54:28 UTC) to January 18, 2026 (23:59:38 UTC), with an average sampling interval of approximately 30 s. The system resource dataset encompassed 5,033,104 samples recorded at 1 s intervals over a 58-day period from November 21, 2025 to January 18, 2026, See Table IV in Section III.

The combined datasets represent, to our knowledge, one of the most extensive empirical characterizations of SRT protocol behavior reported in the literature, exceeding prior studies by an order of magnitude in both temporal coverage and sample volume. Data completeness exceeded 99.7% across both datasets, with minimal gaps attributable to brief maintenance windows and system updates. The 58-day overlap period between the 2 datasets (November 21, 2025 to January 18, 2026) enabled robust cross-correlation analysis between system resources and SRT transport performance. The high-resolution system data (1 s intervals) was aggregated to hourly means to synchronize with the SRT metrics for correlation analysis, preserving sufficient temporal resolution while ensuring measurement alignment.

Descriptive statistics for the 5 SRT transport metrics are summarized in Table V. NAK_rate exhibited a mean of 0.000472 (SD = 0.005126), with values ranging from 0.00001 to 0.383. The distribution was markedly right-skewed, with the bulk of samples clustered near zero and occasional high-magnitude events pulling the tail. RTT_avg displayed a mean of 3.060 ms (SD = 3.047 ms), ranging from 0.130 ms to 121.30 ms—predominantly low-latency transmission punctuated by sporadic spikes. RTT_std, serving as a jitter indicator, averaged 18.695 ms (SD = 14.407 ms) with a maximum of 371.56 ms observed during network instability periods. Buffer variance showed substantial spread (Mean = 557.06, SD = 1560.52), ranging from 0.005 to 138,057, reflecting the receiver buffer's sensitivity to network fluctuations. The cumulative dropped packet counter averaged 310.51 (SD = 38.04) across the monitoring period.

TABLE V. Hourly Descriptive Statistics for SRT Transport Metrics

| Hour | NAK_Rate | | | RTT_Avg | | | Buffer_Var | | |
|---|---|---|---|---|---|---|---|---|---|
| | Mean (ms) | SD (ms) | N | Mean (ms) | SD (ms) | N | Mean (ms) | SD (ms) | N |
| 0 | 0.000205 | 0.000403 | 9996 | 2.81 | 3.03 | 9996 | 563.9 | 1791.9 | 9996 |
| 1 | 0.000196 | 0.000378 | 9572 | 2.61 | 2.46 | 9572 | 458.8 | 798.7 | 9572 |
| 2 | 0.000201 | 0.000365 | 9230 | 2.59 | 2.87 | 9230 | 777.9 | 5982.3 | 9230 |
| 3 | 0.000182 | 0.000328 | 9026 | 2.64 | 2.31 | 9026 | 450.1 | 564.2 | 9026 |
| 4 | 0.000178 | 0.000331 | 8829 | 2.72 | 2.43 | 8829 | 447.7 | 500.7 | 8829 |
| 5 | 0.000170 | 0.000315 | 8961 | 2.85 | 2.53 | 8961 | 502.8 | 732.2 | 8961 |
| 6 | 0.001290 | 0.014617 | 9737 | 2.82 | 2.26 | 9737 | 471.1 | 494.9 | 9737 |
| 7 | 0.000659 | 0.005351 | 11,563 | 2.93 | 2.97 | 11,563 | 505.4 | 680.2 | 11,563 |
| 8 | 0.000779 | 0.005619 | 11,459 | 3.10 | 2.80 | 11,459 | 536.3 | 619.0 | 11,459 |
| 9 | 0.000888 | 0.010653 | 12,000 | 3.14 | 2.79 | 12,000 | 581.0 | 1261.3 | 12,000 |
| 10 | 0.001253 | 0.008332 | 11,710 | 3.40 | 2.82 | 11,710 | 601.8 | 793.2 | 11,710 |
| 11 | 0.000698 | 0.004976 | 12,267 | 3.47 | 2.85 | 12,267 | 612.0 | 835.3 | 12,267 |
| 12 | 0.000812 | 0.005357 | 12,848 | 3.11 | 2.62 | 12,848 | 528.2 | 634.5 | 12,848 |
| 13 | 0.000388 | 0.001342 | 12,345 | 3.27 | 2.91 | 12,345 | 613.2 | 1086.4 | 12,345 |
| 14 | 0.000314 | 0.000881 | 12,079 | 3.64 | 5.53 | 12,079 | 742.9 | 2250.0 | 12,079 |
| 15 | 0.000267 | 0.000695 | 12,582 | 3.12 | 2.56 | 12,582 | 553.3 | 870.7 | 12,582 |
| 16 | 0.000240 | 0.000578 | 12,312 | 3.15 | 2.78 | 12,312 | 557.9 | 685.5 | 12,312 |
| 17 | 0.000195 | 0.000484 | 13,058 | 3.66 | 4.66 | 13,058 | 729.6 | 2302.5 | 13,058 |
| 18 | 0.000779 | 0.009992 | 11,526 | 3.23 | 2.85 | 11,526 | 551.9 | 1425.3 | 11,526 |
| 19 | 0.000356 | 0.001826 | 10,646 | 3.17 | 2.99 | 10,646 | 552.7 | 919.0 | 10,646 |
| 20 | 0.000253 | 0.000782 | 10,097 | 3.29 | 2.80 | 10,097 | 560.6 | 612.3 | 10,097 |
| 21 | 0.000242 | 0.000613 | 10,427 | 2.96 | 2.48 | 10,427 | 507.6 | 542.4 | 10,427 |
| 22 | 0.000237 | 0.000517 | 10,233 | 2.32 | 2.25 | 10,233 | 392.9 | 638.1 | 10,233 |
| 23 | 0.000233 | 0.000462 | 10,169 | 2.71 | 3.00 | 10,169 | 450.7 | 549.9 | 10,169 |

System resource statistics are presented in Table VI. CPU utilization averaged 1.5207% (SD = 1.0680%) during normal operation, with a median of 1.30% and an IQR of 0.80% to 2.00%—a narrow 1.20% band reflecting predominantly low computational load. The maximum of 97.70% occurred during the December 29 anomaly. RAM held remarkably steady at a mean of 26.4209% (SD = 1.0156%), with an IQR spanning only 25.80% to 27.10%. Network receive throughput averaged 6.4739 Mbps (SD = 2.6318 Mbps) reflecting the incoming stream bitrate, while transmit throughput averaged just 0.0458 Mbps, primarily acknowledgment traffic and protocol overhead. (See Table VI and Fig. 1).

TABLE VI. DESCRIPTIVE STATISTICS FOR SYSTEM RESOURCE METRICS

| Statistics | Cpu_Usage (%) | RAM_Usage (%) | Mbps_Recv (Mbps) | Mbps_Sent (Mbps) |
|---|---|---|---|---|
| **count** | 5,033,104 | 5,033,104 | 5,033,104 | 5,033,104 |
| **mean** | 1.5207 | 26.4209 | 6.4739 | 0.0458 |
| **std** | 1.0680 | 1.0156 | 2.6318 | 0.6680 |
| **min** | 0 | 22.60 | 0 | 0 |
| **25%** | 0.80 | 25.80 | 4.6355 | 0.0304 |
| **50%** | 1.30 | 26.50 | 6.7654 | 0.0380 |
| **75%** | 2.00 | 27.10 | 8.3229 | 0.0478 |
| **max** | 97.70 | 66.30 | 385.0047 | 475.3067 |

Hourly system patterns (Fig. 1) showed modest CPU elevation during Hours 6 and 12–14, broadly mirroring the SRT NAK_rate peaks in Fig. 2. RAM usage remained essentially flat across all hours at approximately 26.5%. Network receive throughput rose slightly during afternoon hours (12:00–16:00), and transmit throughput peaked at Hours 14–15 and 23, possibly reflecting batch acknowledgment processing.

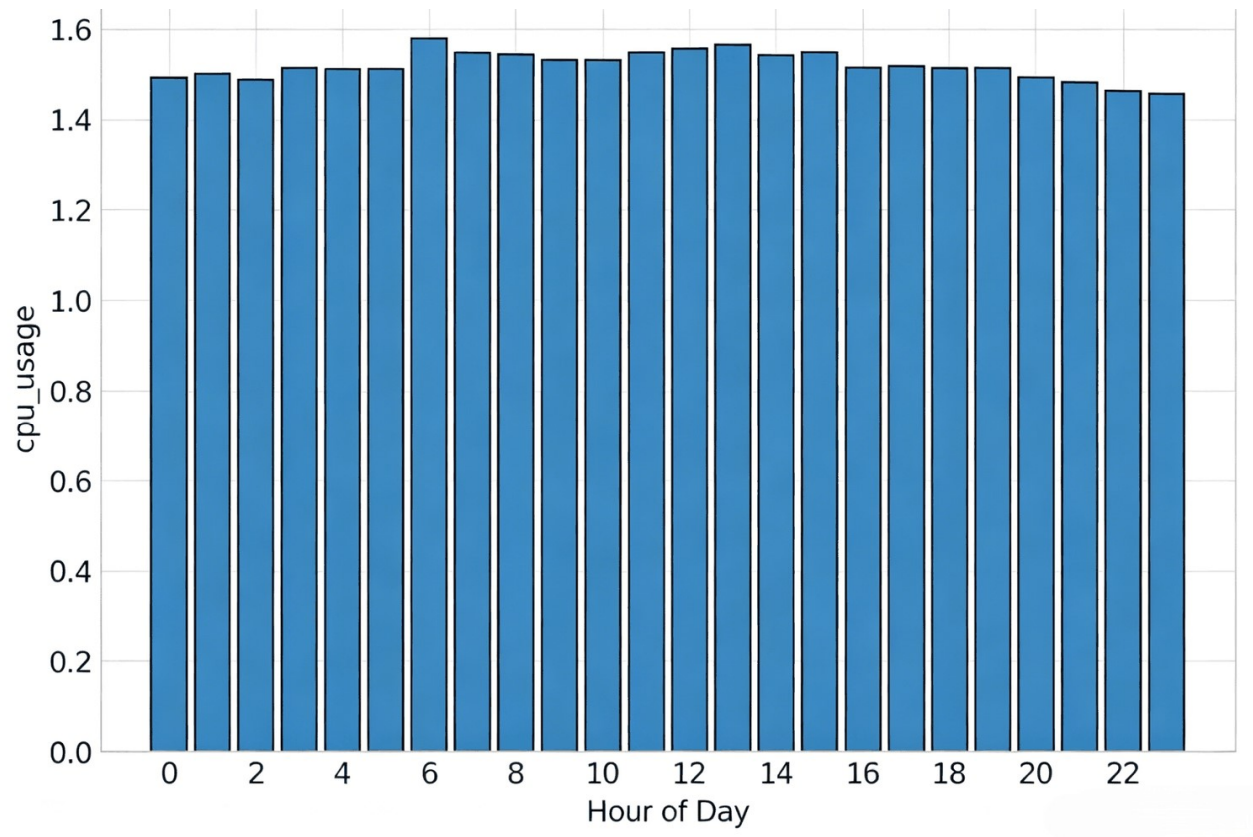

(a)

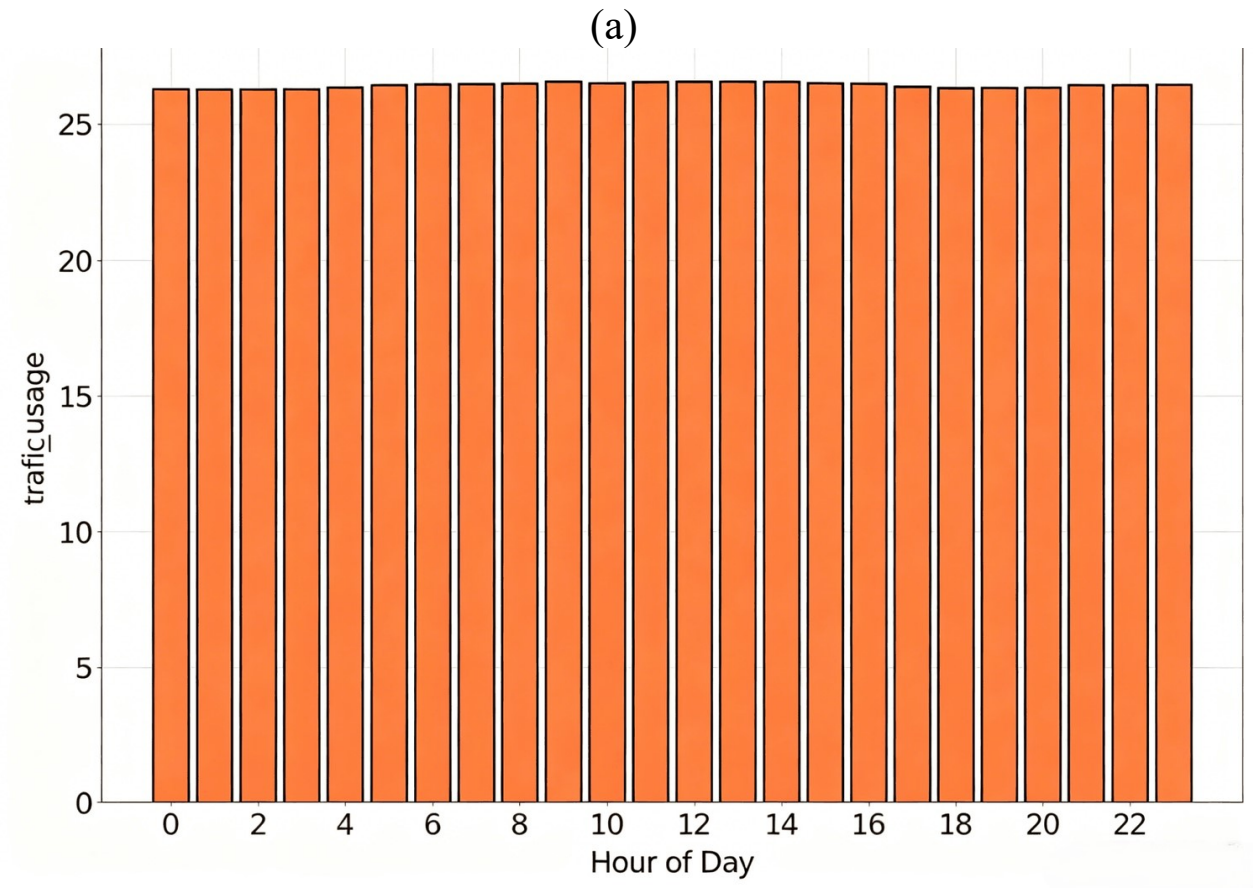

(b)

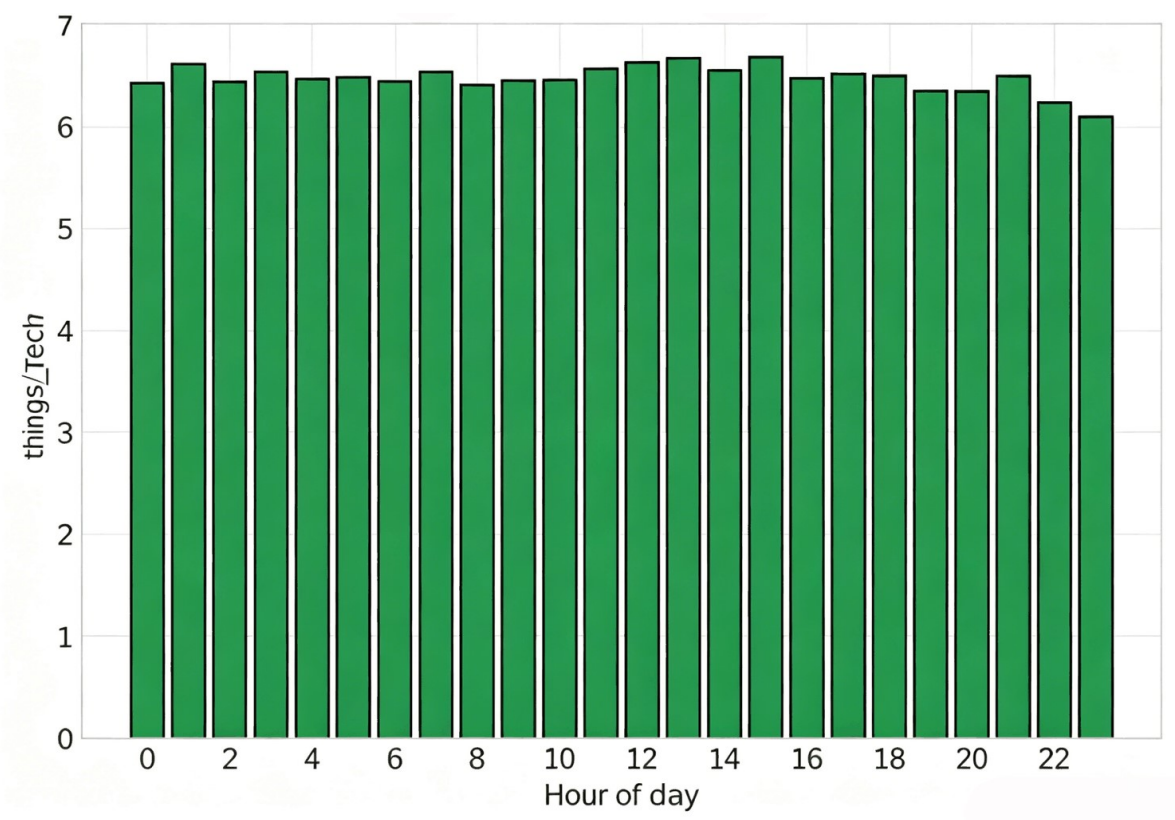

(c)

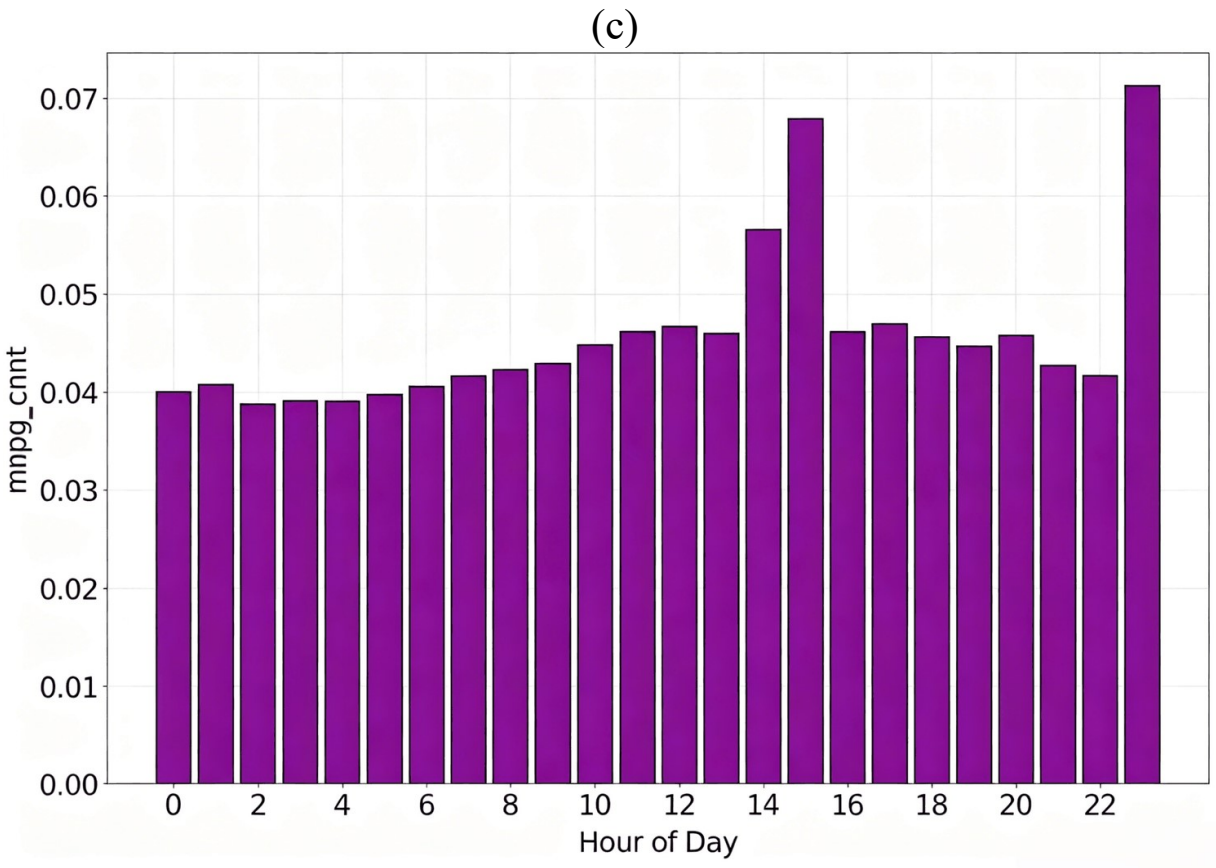

(d)

Fig. 1. Hourly distribution of system resource metrics: (a) CPU Usage by hour, (b) RAM Usage by hour, (c) Network receives by hour. (d) Network sends by hour.

### B. *Temporal Patterns (Hourly, Daily, and Weekly Variation)*

Hourly analysis revealed substantial temporal variation in SRT performance. As presented in Table V and illustrated in Fig. 2, NAK_rate peaked sharply at Hour 6 (Mean = 0.00129, SD = 0.0146), a 7.6-fold increase over the minimum at Hour 5 (Mean = 0.00017, SD = 0.0003). A secondary peak appeared at Hour 10 (Mean = 0.00125, SD = 0.0083), coinciding with mid-morning business activity. The early morning spike at Hour 6 (06:00 local time) likely reflects the transition from overnight quiescence to daytime operations, when the initial surge of daily traffic competes for network resources.

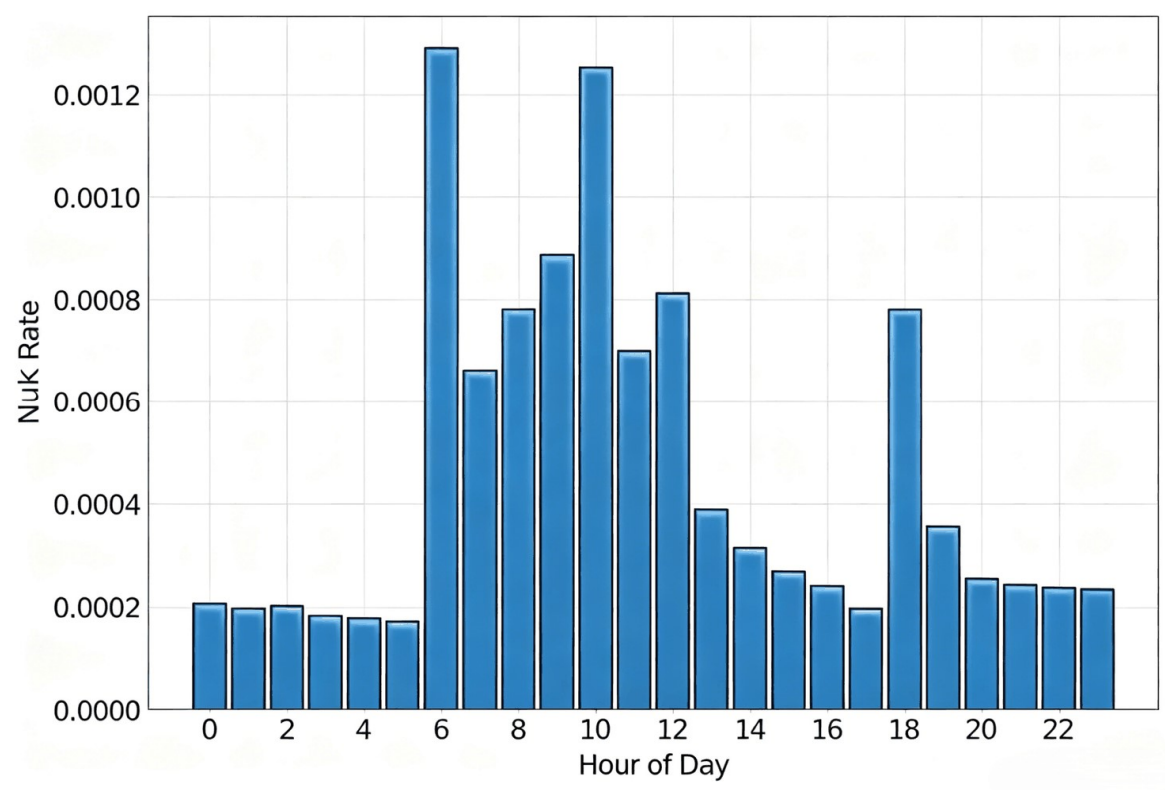

(a)

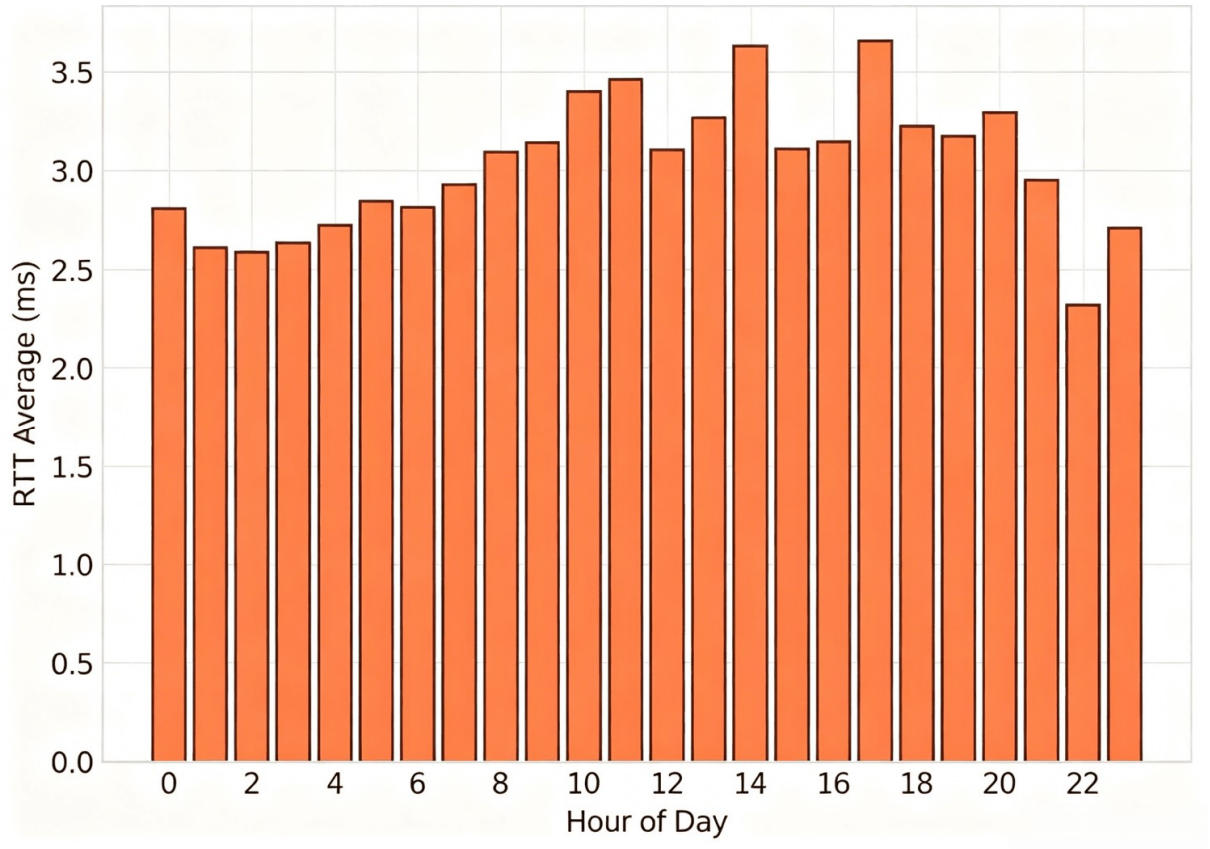

(b)

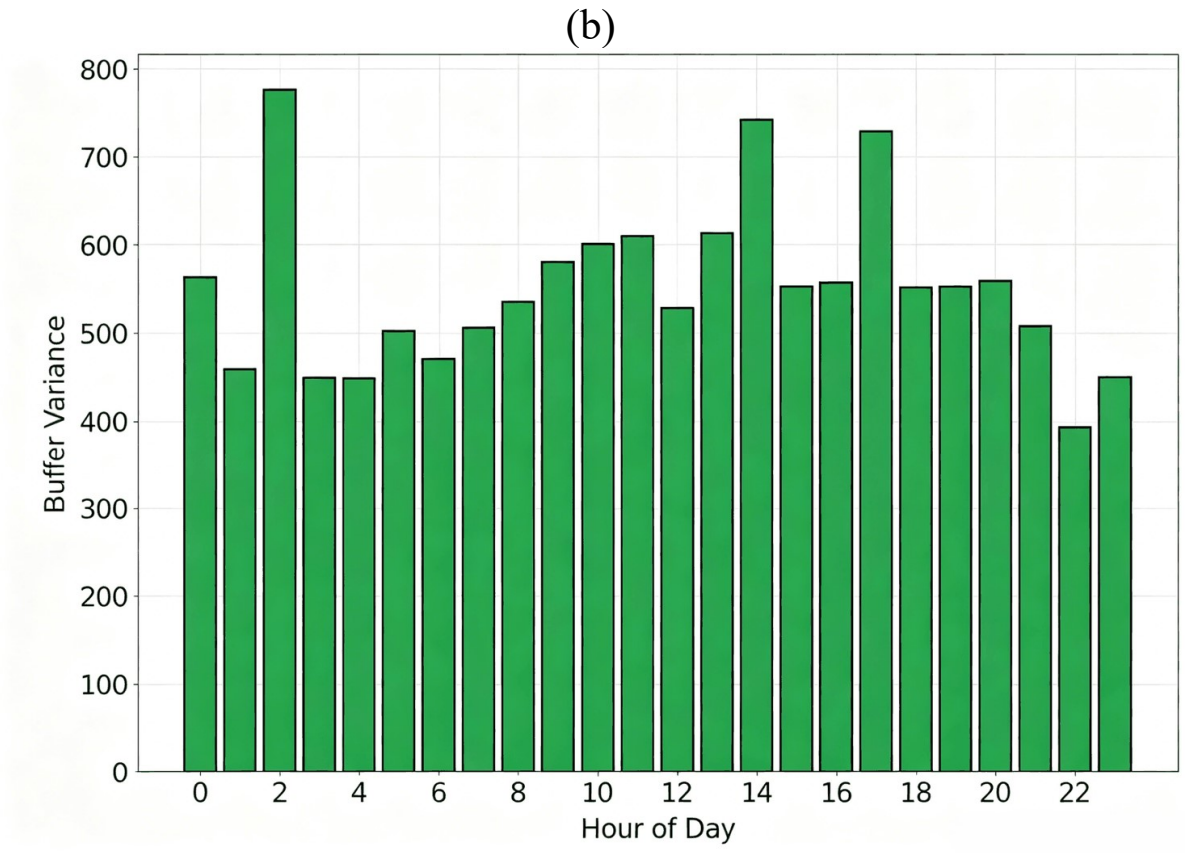

(c)

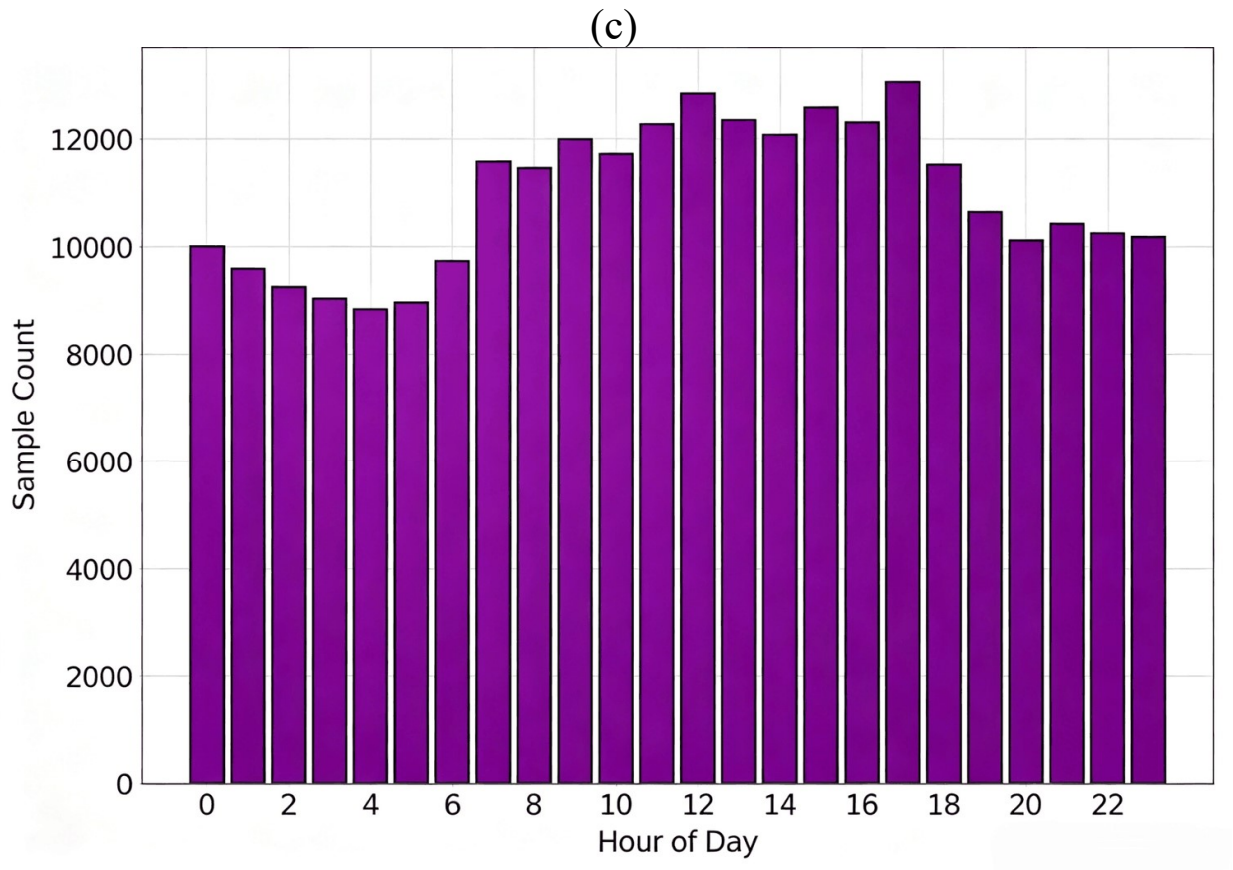

(d)

Fig. 2. Hourly distribution of SRT transport metrics: (a) NAK rate by hour, (b) RTT average by hour, (c) Buffer variance by hour, (d) Sample distribution by hour.

Sample counts across hours ranged from 8829 (Hour 4) to 13,058 (Hour 17), indicating roughly 48% higher streaming activity at midday compared to early morning. RTT_avg showed afternoon elevation, peaking at Hour 14 (Mean = 3.64 ms) and Hour 17 (Mean = 3.66 ms) against a minimum at Hour 2 (Mean = 2.589 ms)—a 40% latency increase during peak hours. Buffer variance followed a bimodal pattern with peaks at Hour 2 (777.89) and Hour 14 (742.90), pointing to distinct sources of buffer instability overnight and in the afternoon.

All hourly variations were statistically significant: ANOVA yielded F = 49.49 ($p$ = 1.69 × $10^{-225}$) for NAK_rate, F = 138.63 ($p < 0.001$) for RTT_avg, and F = 39.88 ($p$ = 1.12 × $10^{-178}$) for buffer_variance, as detailed in Table. The extremely low $p$-values reject the null hypothesis that hourly means are equal, confirming significant diurnal patterns across all measured SRT transport metrics.

Day-of-week effects were equally pronounced. Tuesday exhibited the highest mean NAK_rate (0.000848, SD = 0.007373), while Monday showed the lowest (0.0000736, SD = 0.0000641)—an 11.5-fold gap between the worst and best performing days. Saturday came in third-lowest for NAK_rate (Mean = 0.000143) and lowest for RTT_avg (Mean = 2.663 ms), indicating near-optimal transmission conditions on weekends. (See Table VII and Fig. 3).

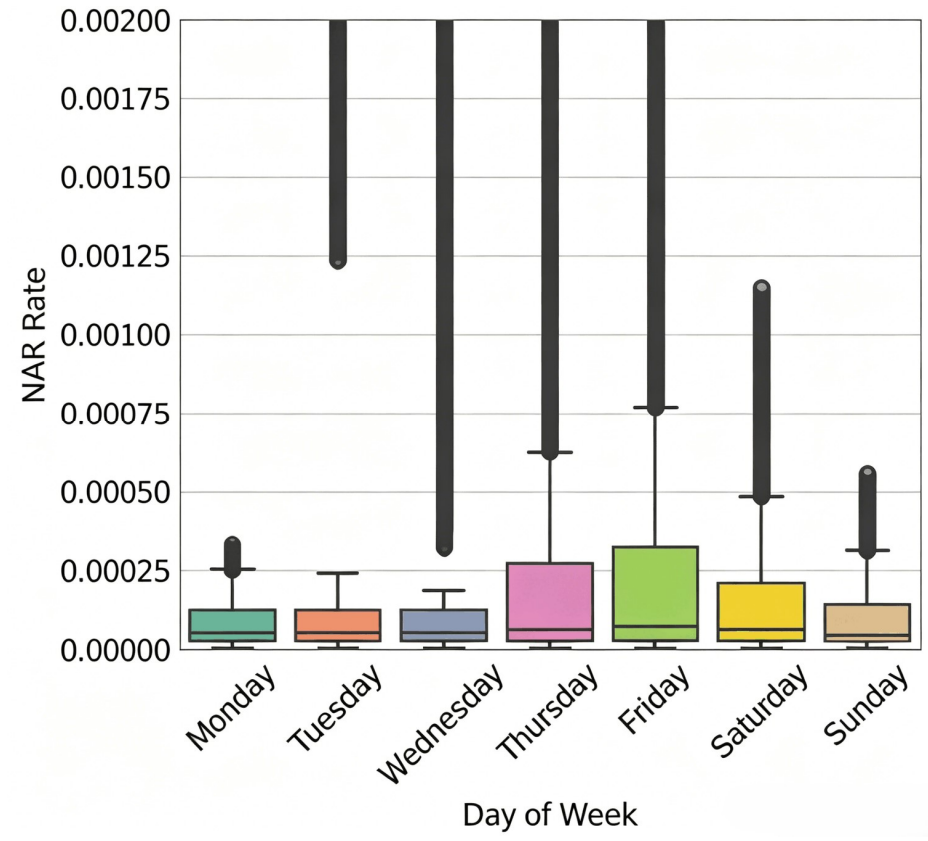

(a)

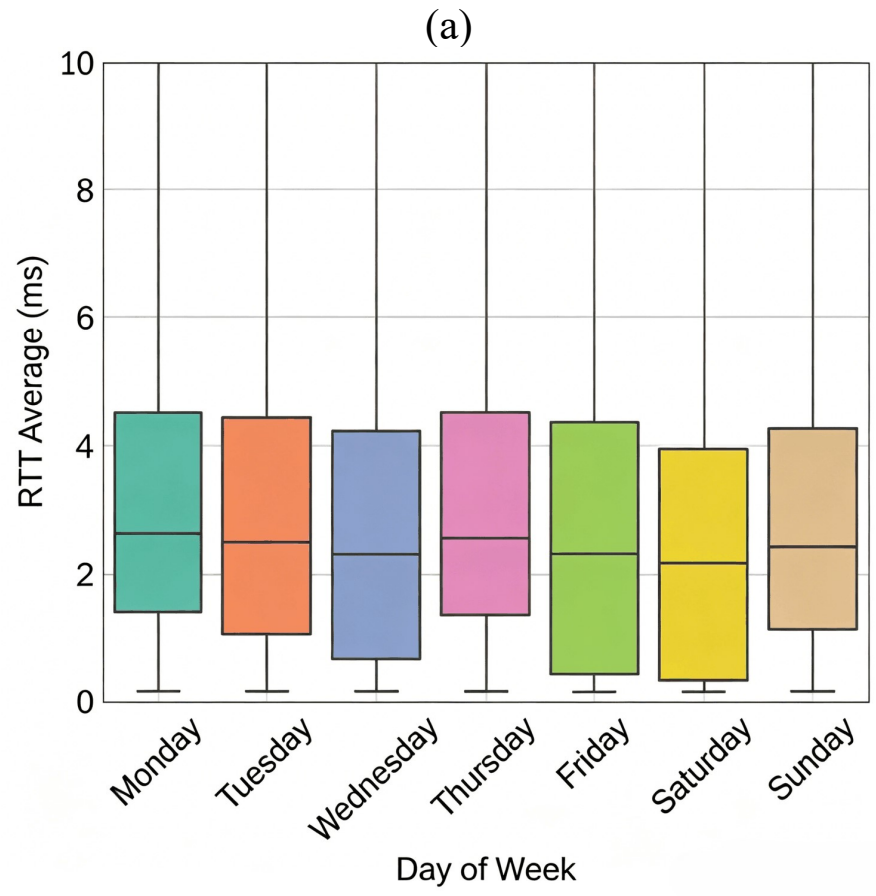

(b)

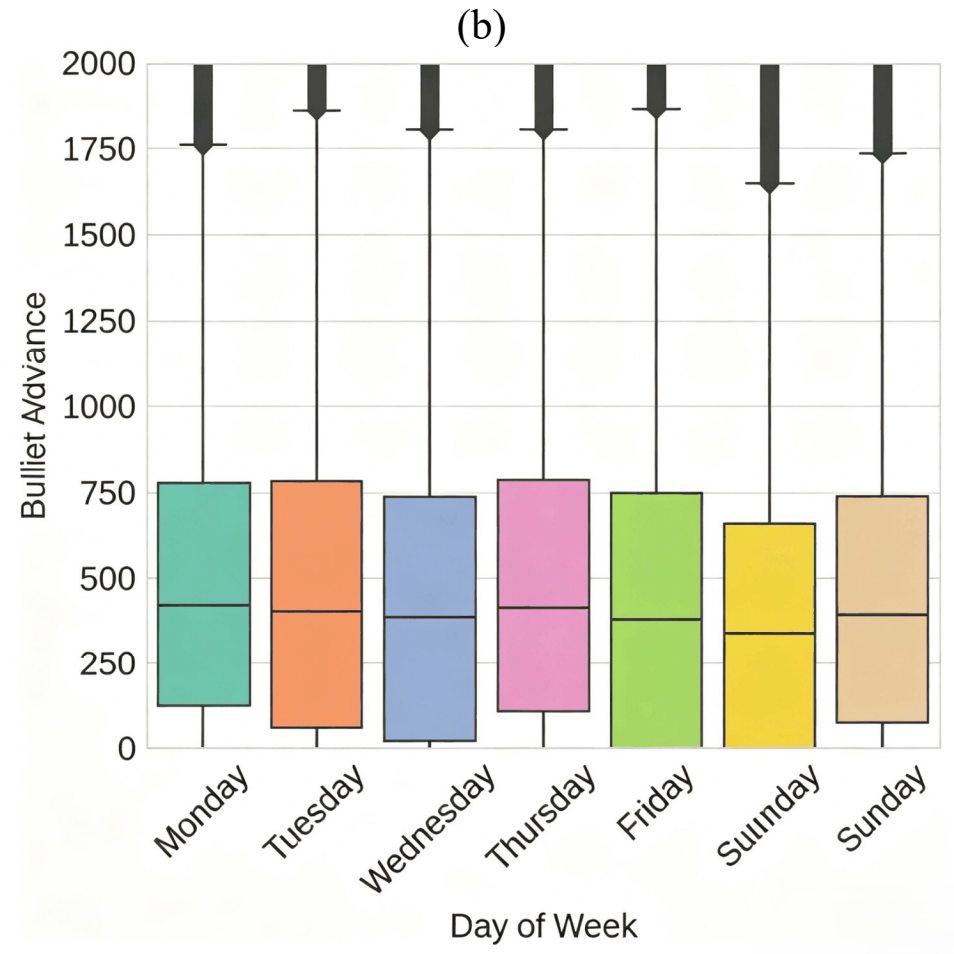

(c)

Fig. 3. Box plot distribution of SRT metrics by day of week: (a) NAK rate by day, (b) RTT by day, (c) buffer variance by day.

TABLE VII. DAILY DESCRIPTIVE STATISTICS BY DAY OF WEEK

| Day | NAK_rate | | RTT_avg | | buffer_variance | |
|---|---|---|---|---|---|---|
| | Mean (ms) | SD (ms) | Mean(ms) | SD(ms) | Mean(ms) | SD(ms) |
| **Monday** | (0.000074) | (0.000064) | 3.2868 | 3.8794 | 575.92 | 1232.79 |
| **Tuesday** | 0.000848 | 0.007373 | 3.1061 | 2.5828 | 526.85 | 606.14 |
| **Wednesday** | 0.000747 | 0.006801 | 2.9925 | 3.3017 | 555.86 | 1414.72 |
| **Thursday** | 0.000703 | 0.007265 | 3.2866 | 2.9557 | 625.78 | 1421.88 |
| **Friday** | 0.000599 | 0.004702 | 3.0579 | 3.1480 | 630.05 | 3050.80 |
| **Saturday** | 0.000143 | 0.000174 | 2.6627 | 2.5017 | 448.76 | 580.92 |
| **Sunday** | (0.000089) | (0.000095) | 3.0223 | 2.7141 | 524.43 | 1045.22 |

Monday and Thursday shared the highest RTT_avg (Mean = 3.287 ms each), while Friday showed the greatest buffer_variance (Mean = 630.05, SD = 3050.80), suggesting increased playout instability toward the end of the workweek. The box plots in Fig. 3 reveal heavy outlier activity on weekdays—particularly Tuesday through Thursday—and comparatively tight distributions on Saturday and Sunday, indicating not only lower means but also reduced variability.

The Kruskal-Wallis H-test confirmed significant differences across days for all metrics: NAK_rate ($H = 6254.67$, $p < 0.001$), RTT_avg ($H = 1397.40$, $p = 8.87 \times 10^{-299}$), and buffer_variance ($H = 1311.69$, $p = 3.20 \times 10^{-280}$), as shown in Table VIII. H-statistics exceeding 1300 indicate substantial between-group variance, confirming that day-of-week is significantly associated with SRT transport variation.

TABLE VIII. STATISTICAL TEST RESULTS FOR TEMPORAL PATTERN ANALYSIS

| Metric | Test | Statistic | *p*-value | Sig. |
|---|---|---|---|---|
| **NAK_rate** | ANOVA (Hourly) | 49.485 | <0.001 | Yes |
| **NAK_rate** | Kruskal-Wallis (Daily) | 6254.667 | <0.001 | Yes |
| **NAK_rate** | Mann-Whitney (Weekend) | $6.97 \times 10^9$ | <0.001 | Yes |
| **RTT_avg** | ANOVA (Hourly) | 138.628 | <0.001 | Yes |
| **RTT_avg** | Kruskal-Wallis (Daily) | 1397.396 | <0.001 | Yes |
| **RTT_avg** | Mann-Whitney (Weekend) | $7.23 \times 10^9$ | <0.001 | Yes |
| **buffer_variance** | ANOVA (Hourly) | 39.877 | <0.001 | Yes |
| **buffer_variance** | Kruskal-Wallis (Daily) | 1311.687 | <0.001 | Yes |
| **buffer_variance** | Mann-Whitney (Weekend) | $7.23 \times 10^9$ | <0.001 | Yes |

A direct weekend-versus-weekday comparison confirmed systematic differences across all metrics. Weekday NAK_rate (Mean = 0.000594) exceeded the weekend value (Mean = 0.000116) by a factor of 5.1. Weekday RTT_avg (Mean = 3.146 ms) ran 10.7% higher than weekend RTT_avg (Mean = 2.843 ms), and weekday buffer_variance (Mean = 582.89) surpassed weekend levels (Mean = 486.60) by 19.8%.

Mann-Whitney U-tests confirmed statistical significance for every comparison: NAK_rate ($U = 6.97 \times 10^9$, $p = 8.78 \times 10^{-12}$), RTT_avg ($U = 7.23 \times 10^9$, $p = 1.08 \times 10^{-110}$), and buffer_variance ($U = 7.23 \times 10^9$, $p = 4.12 \times 10^{-107}$). Weekend network conditions consistently outperformed weekday conditions across every dimension of SRT transport quality.

Heatmaps in Fig. 4 expose the interaction between day and hour. The NAK_rate heatmap Fig. 4(a) reveals a prominent hotspot on Tuesday mornings (Hours 6–10), with values exceeding 0.004 during peak degradation. This Tuesday-morning concentration accounts for a disproportionate share of packet loss events in the entire dataset. Secondary hotspots appear on Wednesday (Hours 10–12) and Thursday (Hours 7–9), suggesting a recurring pattern of early-week network stress during business startup (See Fig. 4).

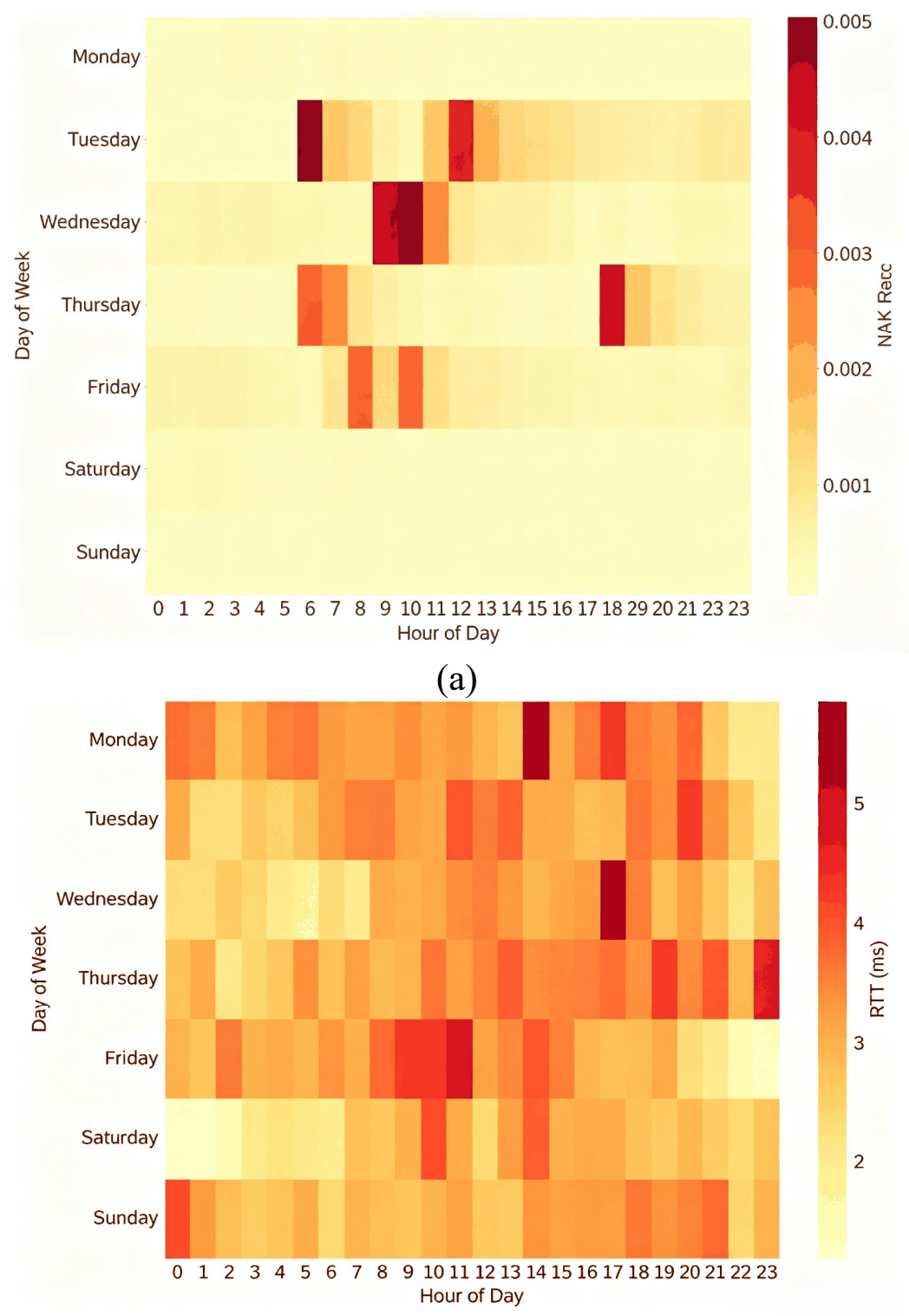

(b)

Fig. 4. Heatmap of (a): NAK_rate and (b): RTT_avg by day and hour.

Saturday and Sunday, by contrast, display uniformly low NAK_rate values (<0.001) at every hour—consistently light-colored rows in the heatmap. This weekend uniformity confirms that the temporal patterns in NAK_rate are primarily driven by workweek activity rather than intrinsic time-of-day effects. The RTT_avg heatmap (Fig. 4(b)) shows elevated latency during afternoon hours (14:00–18:00) on most weekdays, with Monday exhibiting particularly pronounced peaks exceeding 5 ms.

Taken together, the heatmaps indicate that Tuesday mornings represent the most challenging window for SRT transmission quality, while weekend afternoons offer optimal conditions. These interaction patterns are directly actionable: transmission-critical operations and maintenance windows should, where possible, avoid the identified high-degradation periods.

The long-term evolution of SRT metrics is illustrated in Fig. 5, which presents daily aggregated values with 7-day rolling averages across the full 96-day study period. NAK_rate (Fig. 5(a) showed episodic degradation events—sudden elevations followed by gradual recovery—with notable spikes in mid-October, late November, and January 2026. RTT_avg (Fig. 5(b) displayed cyclical weekly oscillations around a stable baseline of roughly 3 ms, and buffer_variance (Fig. 5(c) followed a similar weekly rhythm with a pronounced spike in early January coinciding with elevated NAK_rate values.

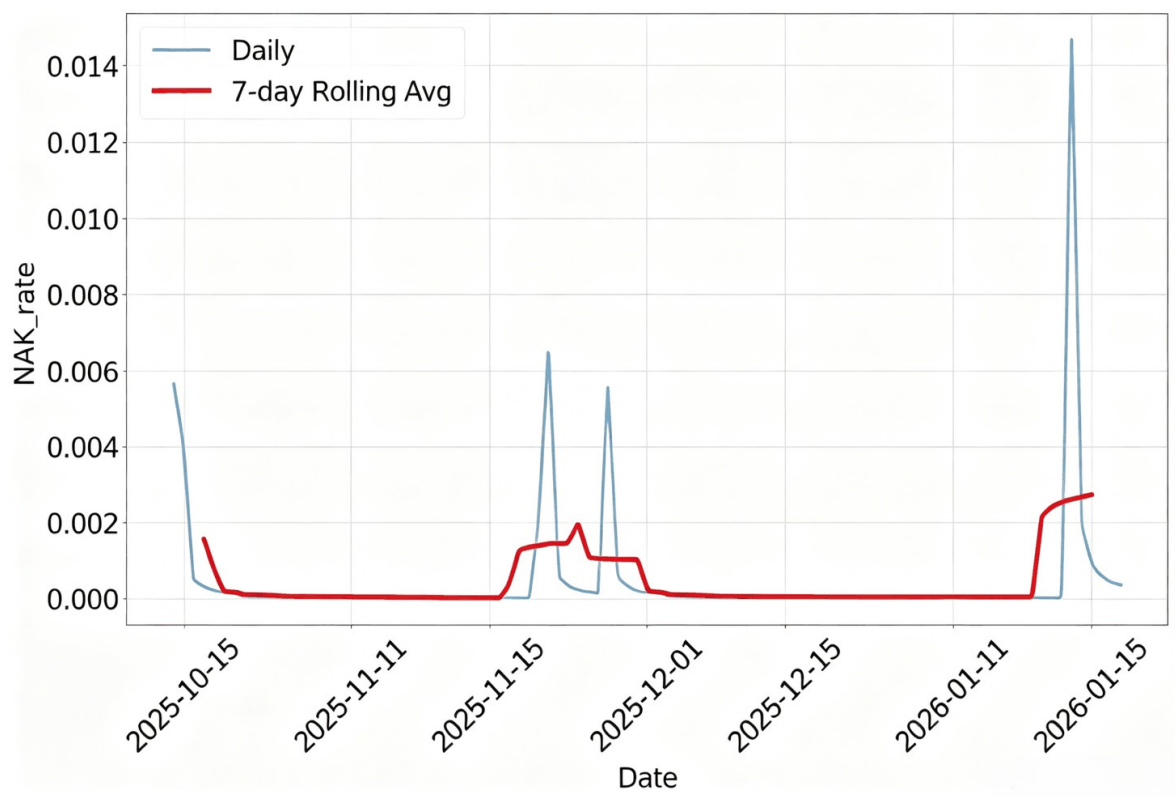

(a)

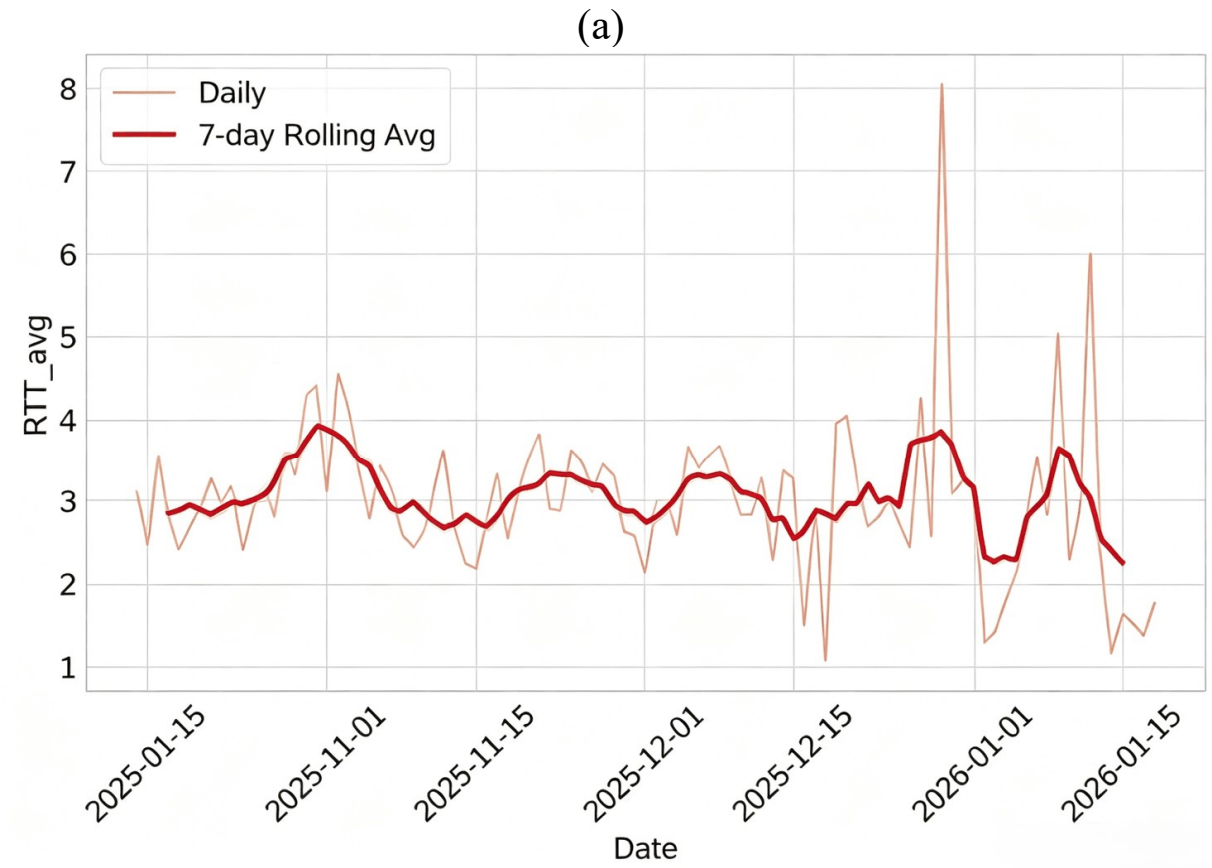

(b)

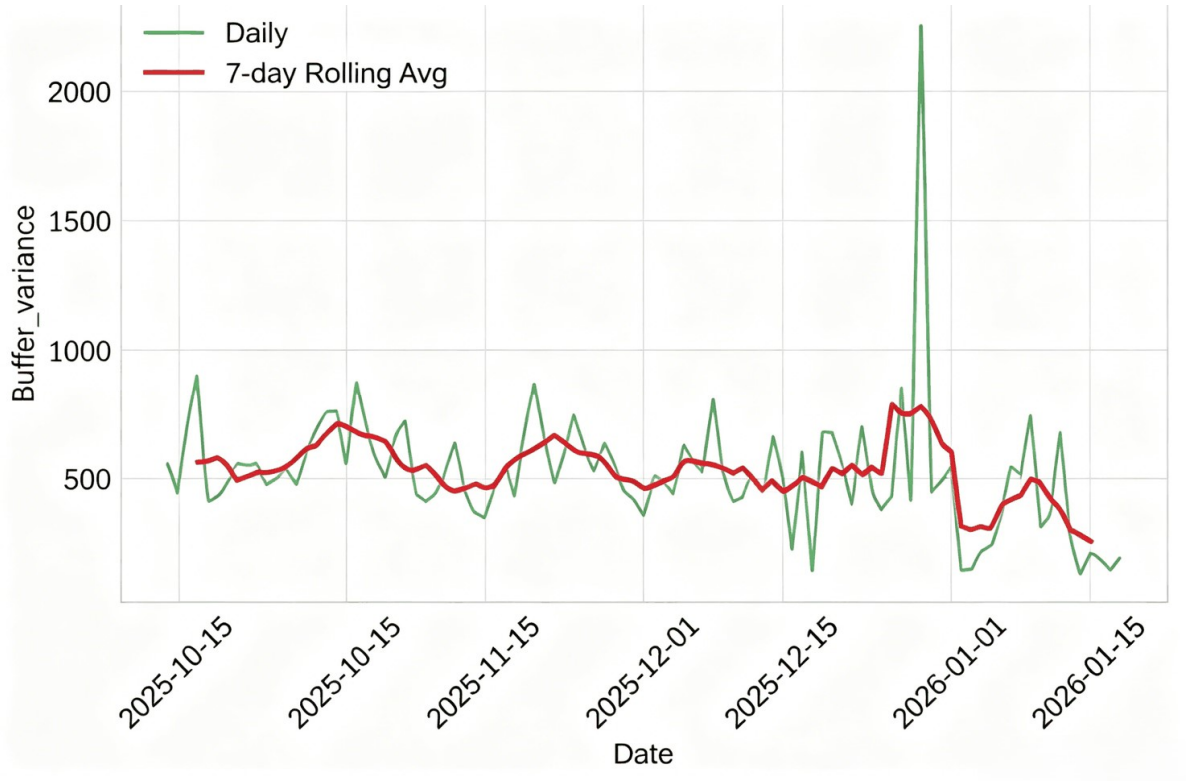

(c)

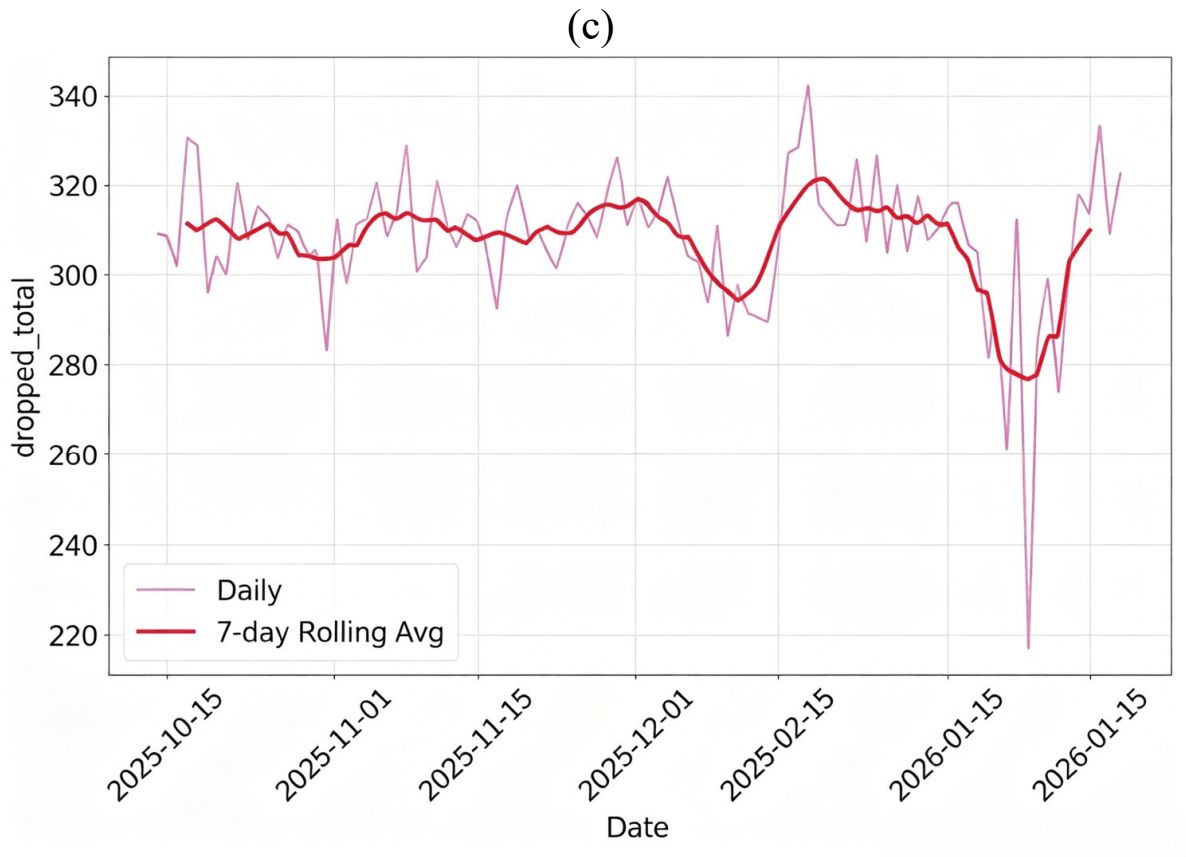

(d)

Fig. 5. 96-day time series with 7-day rolling average: (a) NAK rate, (b) RTT average, (c) Buffer variance, (d) Dropped total.

Classical additive decomposition of the daily NAK_rate series (Fig. 6) separated the data into trend, seasonal, and residual components. The trend component revealed episodic elevations in mid-October, late November, and early January, with extended stability during December. The seasonal component exhibited a clear 7-day period with amplitude of approximately ±0.00075, reinforcing the weekly cyclicality identified in the daily analysis. The residual captured anomalous events, including the early-January spike.

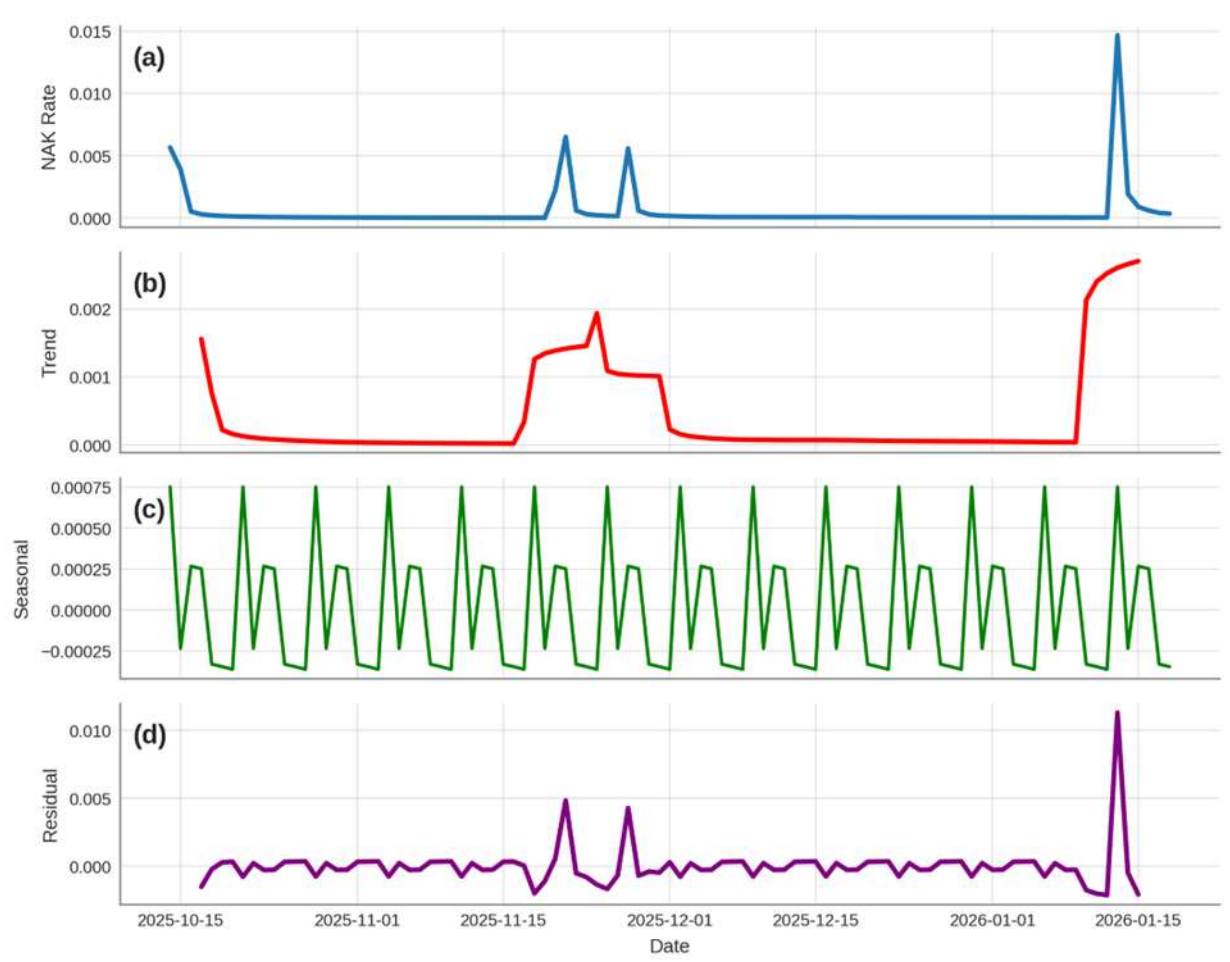

Fig. 6. Additive decomposition of daily NAK_rate: (a) Observed-original NAK rate, (b) Trend component, (c) Seasonal component (weekly pattern), (d) Residual component.

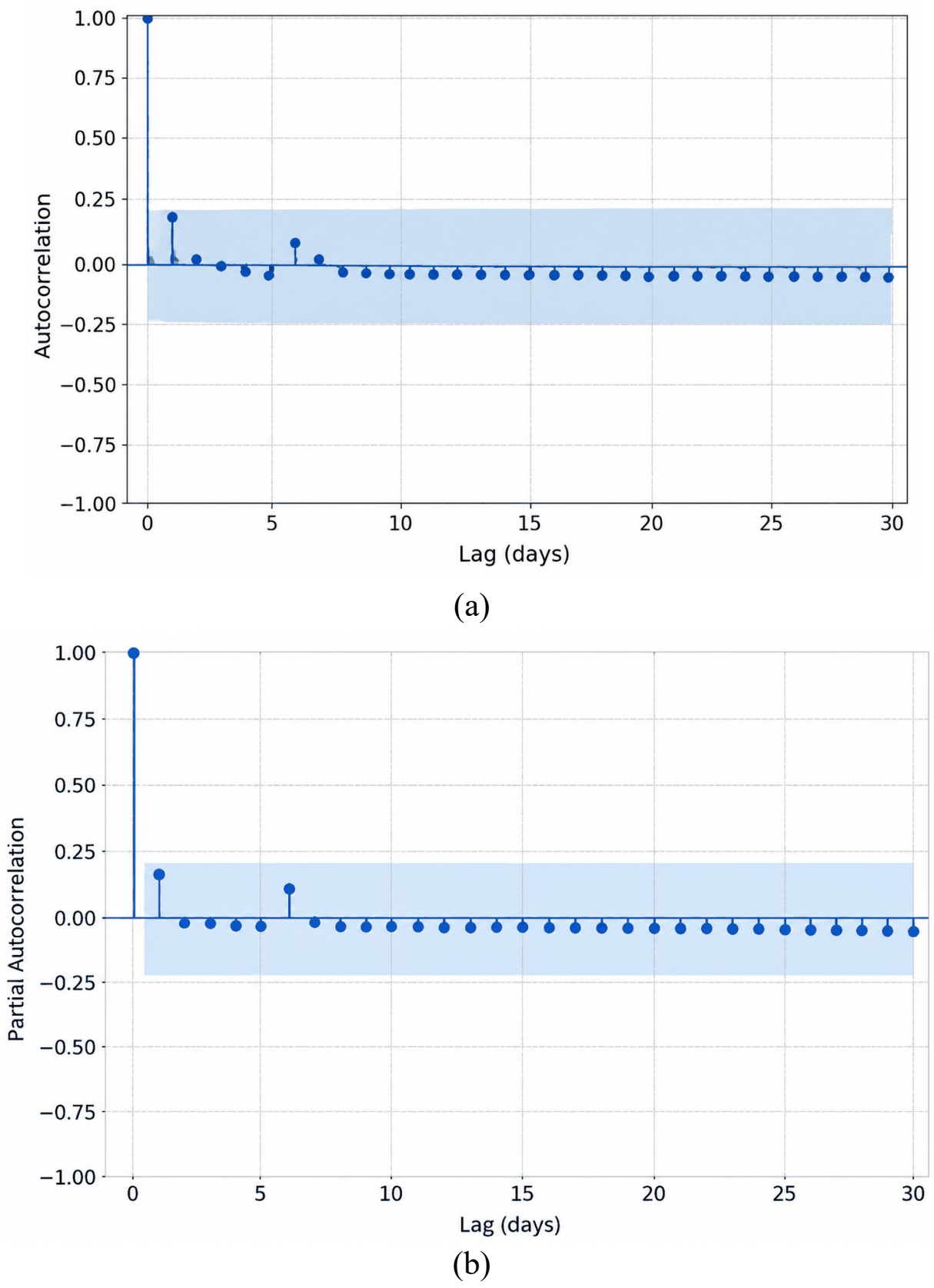

Fig. 7. ACF/PACF: (a) Autocorrelation and (b) Partial Autocorrelation functions.

The ACF of the daily aggregated NAK_rate series Fig. 7(a) showed significant positive autocorrelation at lags 1, 7, and 14 days, all exceeding the 95% confidence bounds. Significance at lags 7 and 14 confirms the weekly periodicity, while lag 1 indicates day-to-day persistence in NAK_rate values (See Fig. 7).

The Partial Autocorrelation Function (PACF) (Fig. 7(b)) revealed significant coefficients at lags 1 and 7 only, with subsequent lags falling within the confidence bounds. This pattern is consistent with an AR (1) process carrying a seasonal AR (7) component: the daily NAK_rate can be partially predicted from the previous day's value and the same day in the prior week. These lag structures inform the parameter selection for SARIMA-based forecasting models in future work.

### C. System-SRT Correlation Analysis

Pearson correlation analysis among SRT metrics (Table IX, Fig. 8) revealed a very strong positive correlation between RTT_avg and RTT_std ($r$ = 0.891), confirming that high-latency periods consistently coincide with elevated jitter. Moderate correlations between RTT_std and buffer_variance ($r$ = 0.669) and between RTT_avg and buffer_variance ($r$ = 0.616) suggest that latency variability propagates to receiver buffer instability. (See Table IX and Fig. 8.)

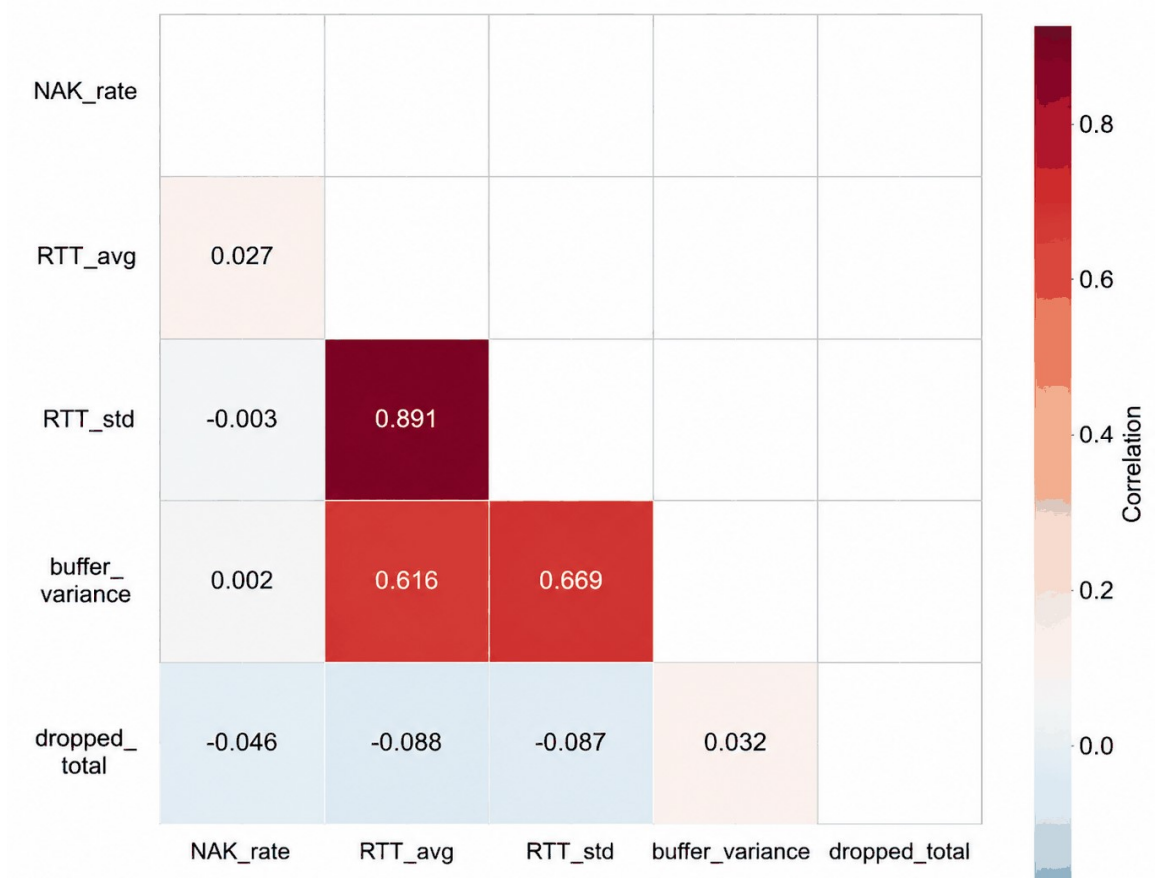

Fig. 8. Correlation heatmap for SRT transport metrics.

TABLE IX. PEARSON CORRELATION MATRIX FOR SRT TRANSPORT METRICS

| Statistics | NAK_rate | RTT_avg | RTT_std | buffer_var | dropped_total |
|---|---|---|---|---|---|
| **NAK_rate** | **1** | 0.027 | −0.003 | 0.002 | −0.046 |
| **RTT_avg** | 0.027 | **1** | 0.891 | 0.616 | −0.088 |
| **RTT_std** | −0.003 | 0.891 | **1** | 0.669 | −0.087 |
| **buffer_var** | 0.002 | 0.616 | 0.669 | **1** | 0.032 |
| **dropped_total** | −0.046 | −0.088 | −0.087 | 0.032 | **1** |

TABLE X. CROSS-CORRELATION MATRIX: SYSTEM RESOURCES VS SRT METRICS

| Statistics | NAK_rate | RTT_avg | buffer_var | cpu_usage | ram_usage | mbps_recv | mbps_sent |
|---|---|---|---|---|---|---|---|
| **NAK_rate** | **1** | 0.279 | −0.004 | 0.617 | 0.449 | 0.094 | −0.121 |
| **RTT_avg** | 0.279 | **1** | 0.636 | 0.493 | 0.509 | 0.359 | 0.264 |
| **buffer_var** | −0.004 | 0.636 | **1** | 0.208 | 0.228 | 0.278 | 0.086 |
| **cpu_usage** | 0.617 | 0.493 | 0.208 | **1** | 0.580 | 0.677 | −0.049 |
| **ram_usage** | 0.449 | 0.509 | 0.228 | 0.580 | **1** | 0.254 | 0.376 |
| **mbps_recv** | 0.094 | 0.359 | 0.278 | 0.677 | 0.254 | **1** | −0.132 |
| **mbps_sent** | −0.121 | 0.264 | 0.086 | −0.049 | 0.376 | −0.132 | **1** |

Note: The differing correlation magnitudes between raw-level and hourly-aggregated analyses reflect a scale-dependent ecological correlation effect (Robinson, 1950) rather than a methodological inconsistency. At finer temporal resolution, high-frequency noise masks underlying relationships; aggregation to hourly means reveals correlations that operate at longer timescales. This phenomenon is examined in detail in Section IV through multi-scale correlation analysis.

NAK_rate, however, showed near-zero correlations with every other metric at the raw 30 s resolution: RTT_avg ($r$ = 0.027), RTT_std ($r$ = −0.003), buffer_variance ($r$ = 0.002), and dropped_total ($r = -0.046$). This indicates that packet loss, as captured by NAK_rate, operates largely as an independent degradation mechanism at short timescales.

Yet when examined at hourly aggregation (Table X), a moderate correlation emerges between NAK_rate and RTT_avg ($r = 0.279$) and a strong one between NAK_rate and CPU usage ($r = 0.617$). How should this discrepancy be interpreted? Table XI presents the NAK_rate–RTT_avg correlation systematically across 6 temporal resolutions, from the native 30 s sampling to daily aggregation.

TABLE XI. MULTI-SCALE CORRELATION ANALYSIS: NAK_RATE VS. RTT_AVG

| Resolution | N | $r$ | $r^2$ | $p$-value | 95% CI |
|---|---|---|---|---|---|
| **30 s (raw)** | 219,016 | 0.042 | 0.002 | <0.001 | [0.037, 0.046] |
| **5-min** | 13,339 | 0.098 | 0.010 | <0.001 | [0.081, 0.115] |
| **15-min** | 4475 | 0.229 | 0.053 | <0.001 | [0.200, 0.257] |
| **1 h** | 1123 | 0.076 | 0.006 | 0.011 | [0.017, 0.134] |
| **4 h** | 283 | 0.222 | 0.049 | <0.001 | [0.107, 0.332] |
| **Daily** | 48 | 0.119 | 0.014 | 0.419 | [−0.170, 0.389] |

At 30 s resolution, the 2 metrics are nearly independent ($r = 0.042$). This is consistent with how SRT's ARQ mechanism works: NAK packets are triggered by sequence-number gaps rather than by RTT values, so instantaneous retransmission demand and propagation delay reflect different physical processes. As the aggregation window widens, the correlation strengthens—peaking at $r = 0.229$ for 15 min—because both metrics share common drivers at the network-load timescale. This ecological correlation effect [37] demonstrates that SRT metric relationships are fundamentally multi-scale: packet-level dynamics exhibit different statistical properties than hour-level trends, and both provide valid but complementary perspectives on protocol behavior.

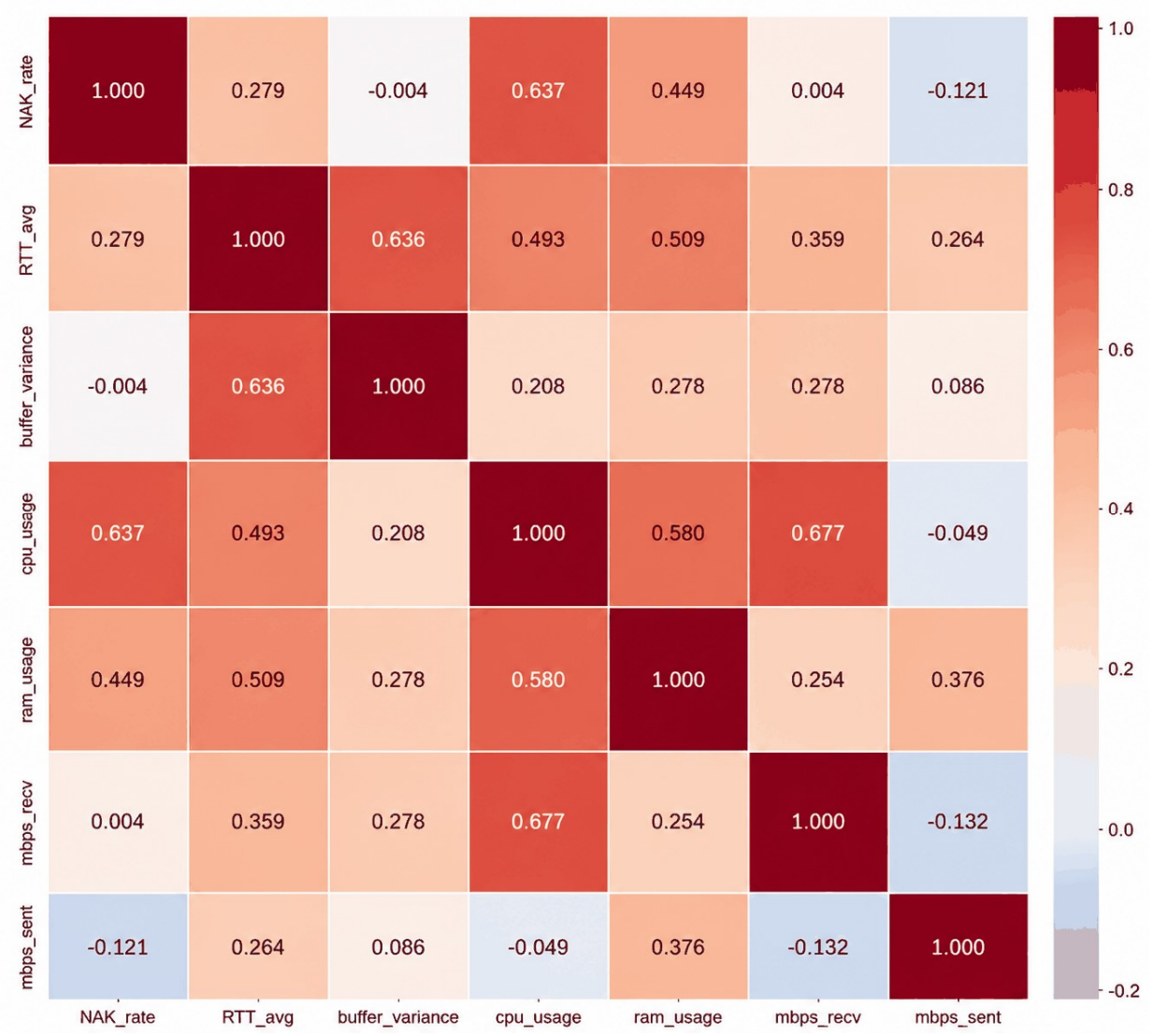

Fig. 9. Cross-correlation heatmap: System vs SRT metrics.

Cross-correlation analysis (Table X, Fig. 9) revealed that the strongest association was between cpu_usage and NAK_rate ($r = 0.617$), indicating that elevated CPU load is associated with increased packet retransmission. CPU also correlated strongly with mbps_recv ($r = 0.677$), reflecting the computational cost of processing incoming network traffic. (See Table X and Fig. 9).

RAM usage showed a moderate correlation with RTT_avg ($r = 0.509$), suggesting that memory pressure is associated with latency elevation. The RAM–NAK_rate correlation ($r = 0.449$) was comparatively weaker, and network receive throughput showed only a weak link with NAK_rate ($r = 0.094$), indicating that raw bandwidth utilization alone does not reliably track packet loss. Overall, these cross-correlations point to CPU utilization as the strongest system-level correlate of SRT transport variation.

### *D. Controlled Attack Impact and Anomaly Detection*

Results from the 6 controlled attack sessions on December 28, 2025 are presented in Table XII and visualized in Fig. 10. 3 low-intensity SRT UDP floods at 500 pps (Attacks 1–3) and three medium-intensity UDP floods at ~1000 pps (Attacks 4–6) were executed between 20:10 and 21:07 local time. The pre-attack baseline showed mean CPU of 1.81%, RAM at 25.24%, and network throughput of 8.08 Mbps. Detailed impact measurements are provided in Table XIII. Low-intensity floods produced CPU increases of +176% to +806%, with peak values between 5.0% and 16.4%. Attack 3 showed notably reduced impact (5.0% peak versus 15%–16% for Attacks 1–2); possible explanations include OS-level caching of connection handling routines or natural variation in baseline activity. Medium-intensity floods hit harder, with CPU spikes of +662% to +1348%. Attack 4 drove CPU to 26.2%, the most severe attack-induced response observed.

Throughout all 6 sessions, Random Access Memory (RAM) remained stable at 25%–26%, confirming that the attack traffic primarily consumed computational rather than memory resources. Network receive throughput showed minimal change because the attack traffic was generated locally within the isolated laboratory segment.

A natural follow-up question is whether the dramatic CPU spikes translated into visible degradation at the SRT transport layer. To answer this, SRT metrics (NAK_rate, RTT_avg, buffer_variance, dropped_total) were extracted for each attack window. Because SRT statistics were sampled at roughly 60 s intervals during this period, each attack yielded between 3 and 10 transport-layer data points (see Table XIV).

TABLE XII. COMPARATIVE ANALYSIS OF ATTACK SESSIONS

| Attack | Local Time | Samples | CPU Max | CPU change% | RAM |
|---|---|---|---|---|---|
| Attack 1–SRT Low | 20:10–20:15 | 298 | 15.4 | 751 | 25.88 |
| Attack 2–SRT Low | 20:16–20:22 | 357 | 16.4 | 806 | 26.22 |
| Attack 3–SRT Low | 20:22–20:27 | 298 | 5.0 | 176 | 25.32 |
| Attack 4–UDP Med | 20:30–20:38 | 477 | 26.2 | 1348 | 26.34 |
| Attack 5–UDP Med | 20:40–20:50 | 596 | 14.4 | 696 | 25.78 |
| Attack 6–UDP Med | 20:57–21:07 | 596 | 13.8 | 662 | 26.39 |

TABLE XIII. CONTROLLED ATTACK BASELINE AND IMPACT SUMMARY (DECEMBER 28, 2025)

| Period | Time (Local) | CPU Mean (%) | CPU Max (%) | RAM Mean (%) |
|---|---|---|---|---|
| **Pre-Attack Baseline** | 19:40–20:10 | 1.81 | 4.2 | 25.24 |
| **Attack Period** | 20:10–21:07 | 2.15 | 26.2 | 25.99 |
| **Post-Attack Recovery** | 21:07–21:30 | 1.76 | 3.8 | 25.31 |

TABLE XIV. SRT TRANSPORT METRICS DURING CONTROLLED ATTACK SESSIONS

| Period | NAK_rate | Δ% | RTT (ms) | Δ% | Buf.Var | Dropped | CPU Max | N_srt |
|---|---|---|---|---|---|---|---|---|
| **Baseline** | **0.000054** | — | **2.718** | — | **330.9** | **311** | **1.86%** | **82** |
| **Atk 1 (SRT 500)** | 0.000047 | −13% | 1.000 | −63% | 64.4 | 273 | 15.4% | 7 |
| **Atk 2 (SRT 500)** | 0.000049 | −9% | 2.309 | −15% | 262.1 | 285 | 16.4% | 7 |
| **Atk 3 (SRT 500)** | 0.000051 | −4% | 3.521 | +30% | 416.4 | 299 | 5.0% | 3 |
| **Atk 4 (UDP 1K)** | 0.000051 | −5% | 3.486 | +28% | 360.2 | 296 | 26.2% | 10 |
| **Atk 5 (UDP 1K)** | 0.000055 | +3% | 0.479 | −82% | 22.7 | 322 | 14.4% | 6 |
| **Atk 6 (UDP 1K)** | 0.000056 | +4% | 3.413 | +26% | 479.6 | 324 | 13.8% | 6 |
| **Recovery** | **0.000055** | **+2%** | **1.798** | — | **190.0** | **320** | — | **30** |

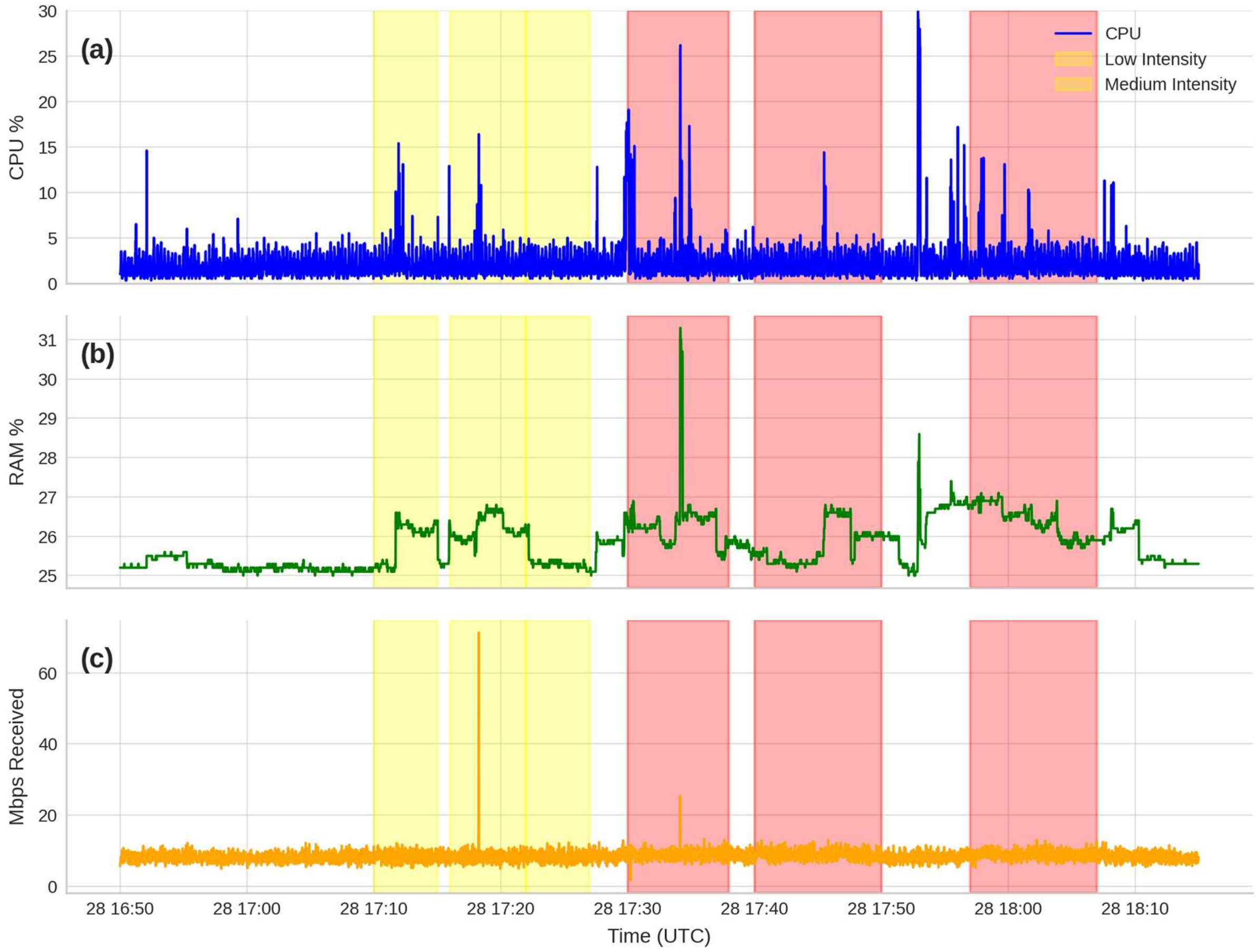

Fig. 10. System metrics during controlled attacks (Dec. 28): (a) CPU usage, (b) RAM usage, (c) Network receive throughput.

Before examining the results, a note on the chosen attack intensities is warranted. The monitored SRT stream operated at a mean bitrate of 5.62 Mbps, corresponding to approximately 517 packets per second with standard 1316-byte SRT payloads and 42-byte UDP/IP overhead. The 500 pps attack therefore matched the legitimate stream rate almost exactly (0.97×), while 1000 pps represented roughly twice the stream rate (1.93×). These ratios align with the low-rate DDoS literature, where attack traffic at or below the legitimate rate can cause protocol-level disruption while evading volume-based detection [30].

The table reveals a striking asymmetry. While CPU surged by 176%–1348% above baseline, SRT transport metrics barely moved: NAK_rate changes stayed within −13% to +4% of pre-attack values, and the dropped_total counter shifted by only −39 to +13 packets. After the final attack, NAK_rate recovered to within 2.3% of baseline within minutes.

Aggregating across attack types, the 3 SRT Low sessions (500 pps) produced mean CPU_max of 12.3% ± 6.3% with a mean NAK_rate shift of −8.5% ± 4.2%, while the three UDP Medium sessions (1000 pps) produced mean CPU_max of 18.1% ± 7.0% with a NAK_rate shift of +0.6% ± 5.1%. The difference between attack types was not statistically significant for NAK_rate impact ($t = -2.40$, $p = 0.074$), indicating that SRT's ARQ mechanism provided comparable transport-layer resilience against both vectors despite their substantially different CPU footprints.

This finding has a direct security implication: monitoring SRT transport statistics alone would not have flagged any of the 6 attacks. The protocol's built-in error recovery effectively masks the system-level impact. Reliable intrusion detection for SRT infrastructure therefore requires multi-metric monitoring that includes system resources—particularly CPU—alongside transport indicators.

An unexpected event on December 29, 2025—one day after the controlled experiments—provided an unplanned test of the detection methodology. At 07:07:56 UTC, CPU spiked to 97.70%, the highest value in the entire 96-day record and a +5328% increase over the preceding hour's mean of 1.80% (06:00–07:00 UTC). This exceeded the worst controlled attack (Attack 4, 26.2%) by a factor of 3.7. See Table XV and Fig. 11.

TABLE XV. UNEXPECTED ANOMALY EVENT METRICS (DECEMBER 29, 2025)

| Period | Time (Local) | CPU mean (%) | CPU max (%) | RAM mean (%) | RAM max (%) |
|---|---|---|---|---|---|
| **Before Anomaly** | 09:50–10:05 | 1.71 | 10.10 | 25.51 | 26.1 |
| **During Anomaly** | 10:05–10:15 | 3.64 | 97.70 | 26.03 | 29.8 |
| **After Anomaly** | 10:15–10:30 | 1.71 | 5.80 | 25.89 | 26.1 |

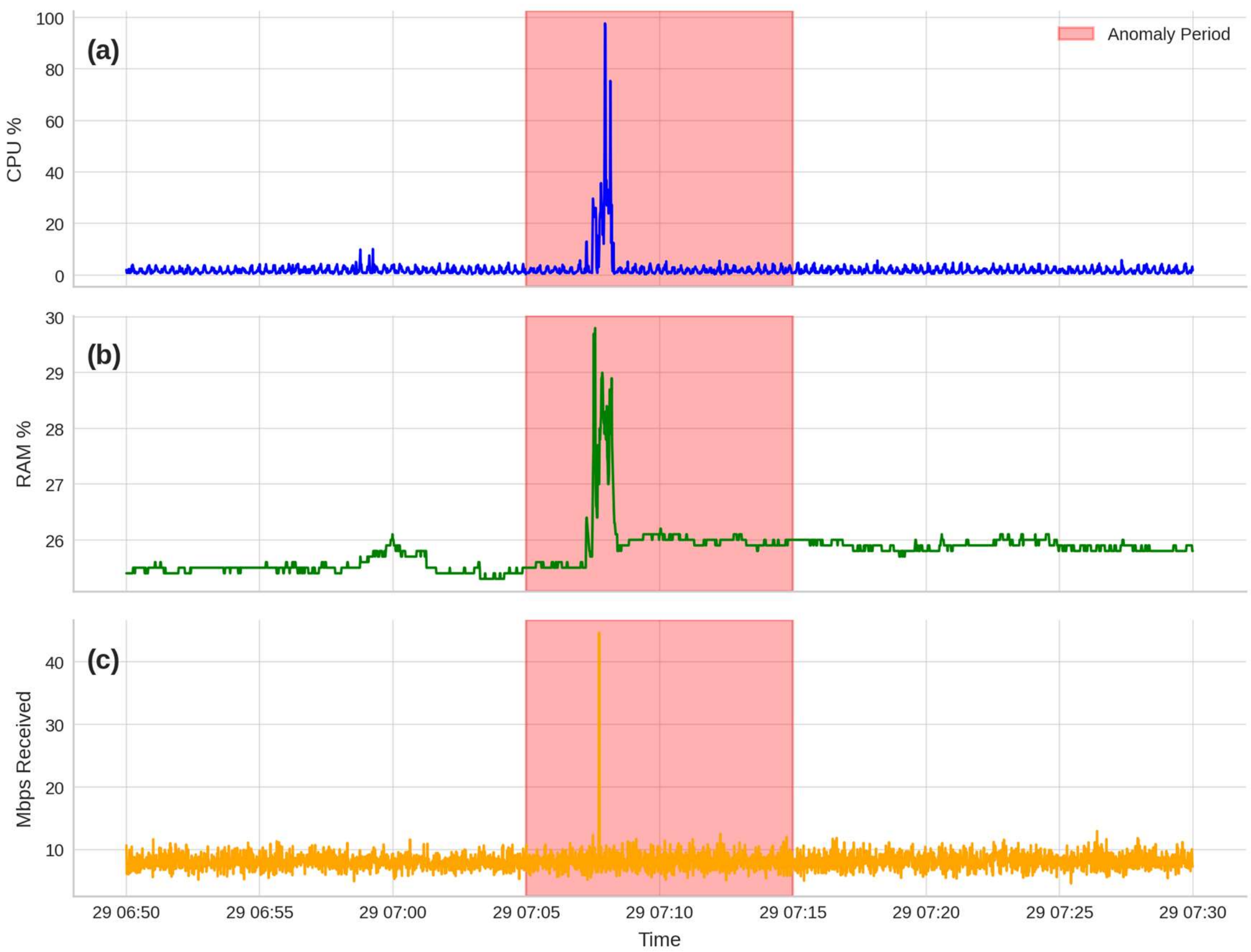

Fig. 11. System metrics during unexpected anomaly (Dec. 29): (a) CPU usage, (b) RAM usage, (c) Network receive throughput.

Network receive throughput peaked within the same two-minute window at 44.66 Mbps (+449% from baseline 8.14 Mbps), while RAM rose modestly to 29.80% (+17%). Recovery was rapid: within 10 min, all metrics returned to baseline levels.

The anomaly's signature differed qualitatively from the controlled attacks. The 6 planned attacks produced CPU-dominant signatures with essentially flat network throughput, because the attack traffic was generated locally. The December 29 event, by contrast, showed simultaneous CPU and network elevation—consistent with an external traffic event, though definitive attribution is not possible without packet capture data. Possible explanations include external traffic ingress, scheduled maintenance or backup operations, automated software updates, or a legitimate traffic surge. Notably, SRT transport metrics during the anomaly remained stable (NAK_rate = 0.000054, within 0.7% of baseline), further demonstrating the protocol's resilience under extreme system-level disruption.

A summary of all 7 security-related events is presented in Table XVI, with the revised source attribution.

TABLE XVI. SECURITY EVENTS SIGNATURE COMPARISON

| Event | Date | Time (Local) | Type | CPU max% | Net Max | Signature | Source |
|---|---|---|---|---|---|---|---|
| **Attack 1** | Dec. 28 | 20:10–20:15 | SRT UDP Low | 15.4 | — | CPU-only | Internal |
| **Attack 2** | Dec. 28 | 20:16–20:22 | SRT UDP Low | 16.4 | — | CPU-only | Internal |
| **Attack 3** | Dec. 28 | 20:22–20:27 | SRT UDP Low | 5.0 | — | CPU-only | Internal |
| **Attack 4** | Dec. 28 | 20:30–20:38 | UDP Medium | 26.2 | — | CPU-only | Internal |
| **Attack 5** | Dec. 28 | 20:40–20:50 | UDP Medium | 14.4 | — | CPU-only | Internal |
| **Attack 6** | Dec. 28 | 20:57–21:07 | UDP Medium | 13.8 | — | CPU-only | Internal |
| **Anomaly** | **Dec. 29** | **10:07:56** | **Unknown** | **97.7** | **44.66** | **CPU+Network** | **Undetermined** |

TABLE XVII. COMPARATIVE ANOMALY DETECTION PERFORMANCE

| Method | Threshold | Detections | FP Rate | Events (7) | Missed | Verdict |
|---|---|---|---|---|---|---|
| **Fixed (>10%)** | >10.0% | 76 | 0.029% | 6/7 | Atk 3 | Inadequate |
| ***Z*-score (3$\sigma$)** | >3.16% | 378 | 0.146% | 7/7 | None | **Recommended** |
| **Moving Avg** | Adaptive | 814 | 0.314% | 7/7 | None | Higher FP |
| **IQR (1.5×)** | >2.63% | 2,090 | 0.806% | 7/7 | None | Highest FP |

TABLE XVIII. EFFECT SIZES FOR TEMPORAL STATISTICAL TESTS ($N$ = 219,016)

| Comparison | Test | Metric | Statistic | $\eta^2/d$ | Class | $p$ | N |
|---|---|---|---|---|---|---|---|
| Hourly (24h) | ANOVA | NAK_rate | $F = 49.49$ | 0.0048 | Negligible | <0.001 | 262,672 |
| Hourly (24h) | K–W | RTT_avg | $H = 4135$ | 0.0188 | Small | <0.001 | 262,672 |
| Daily (DOW) | K–W | NAK_rate | $H = 6255$ | 0.0315 | Small | <0.001 | 262,672 |
| Daily (DOW) | K–W | RTT_avg | $H = 1397$ | 0.0085 | Negligible | <0.001 | 262,672 |
| Wkend/Wkday | M–W | NAK_rate | $d = -0.098$ | — | Negligible | <0.001 | 262,672 |
| Wkend/Wkday | M–W | RTT_avg | $d = -0.111$ | — | Negligible | <0.001 | 262,672 |

Two distinct signatures emerge: (1) CPU-dominant events (all 6 controlled attacks), where CPU rises sharply while network throughput holds steady, consistent with locally-generated traffic; and (2) CPU-plus-network events (the December 29 anomaly), where both metrics spike simultaneously. This differentiation provides a starting point for automated event classification, though packet-level capture would be needed for conclusive attribution.

The fixed 10% CPU threshold used in the initial analysis detected 6 of 7 known events but missed Attack 3 (CPU_max = 5.0%). To evaluate alternative approaches, four detection methods were applied to the 259,201-sample CPU baseline (Mean = 1.41%, SD = 0.58%, skewness = 14.87, kurtosis = 582.95) (see Table XVII).

The *Z*-score method (3$\sigma$ above mean, threshold = 3.16%) detected all seven events with a false positive rate of 0.146%—manageable in a production monitoring context. The moving-average approach also achieved perfect detection but at more than double the false positive rate. The IQR method cast the widest net, flagging over 2000 data points with a false positive rate 5.5× that of the *Z*-score approach.

The failure of the fixed threshold to catch Attack 3 is a consequential finding in its own right. It demonstrates that low-rate SRT-specific attacks can slip beneath static detection criteria calibrated for higher-intensity events. For production deployments, we recommend the *Z*-score method as the primary detection criterion, supplemented by moving-average deviation analysis for identifying gradual degradation trends that statistical thresholds might miss. Whatever method is adopted, thresholds should be derived from site-specific baselines rather than fixed values, and recalibrated periodically as workload profiles evolve.

### *E. Comparative Analysis and Effect Size Interpretation*

The preceding sections established statistical significance for all temporal comparisons. However, given the large sample size ($N$ > 219,000), even trivial differences can reach significance. To gauge whether these patterns carry practical weight, we computed effect sizes for each comparison. Table XVIII presents eta-squared ($\eta^2$) for ANOVA and Kruskal-Wallis tests, and Cohen's d for Mann-Whitney pairwise comparisons.

Effect sizes were uniformly small: $\eta^2$ ranged from 0.005 to 0.032, meaning that temporal period explains at most 3.2% of total metric variance. The weekend/weekday Cohen's d values (−0.098 for NAK_rate, −0.111 for RTT_avg) fall squarely in the "negligible" range. This combination of highly significant $p$-values with small effect sizes is characteristic of large-sample studies and reflects the statistical power of the dataset rather than substantively large group differences.

Where, then, does the practical relevance lie? Two observations stand out. First, the peak hourly NAK_rate of 0.129% during Hour 6 exceeds the 0.1% packet-loss threshold typically applied in professional SD broadcasting. During those hours, the quality margin available for additional network stressors narrows considerably. Second, the overall mean NAK_rate of 0.0468% and mean RTT of 3.107 ms remain well within acceptable limits (RTT < 200 ms for live broadcast), confirming that the deployment maintained adequate quality across the full 96-day period. In short, temporal variation is modest in the aggregate but operationally significant during specific peak windows when metrics approach threshold boundaries (see Table XIX).

TABLE XIX. 95% CONFIDENCE INTERVALS FOR KEY METRICS

| Metric | N | Mean | 95% CI | Source |
|---|---|---|---|---|
| **NAK_rate** | 219,016 | 0.000468 | [0.000447, 0.000489] | SRT raw |
| **RTT_avg (ms)** | 219,016 | 3.107 | [3.095, 3.120] | SRT raw |
| **CPU_usage (%)** | 259,201 | 1.414 | [1.412, 1.416] | System |
| **CPU↔NAK *r*** | 1392 | 0.617 | [0.583, 0.648] | Hourly agg. |
| **NAK↔RTT *r*** | 219,016 | 0.042 | [0.037, 0.046] | SRT raw |

### F. *Discussion: Implications, Limitations, and Future Directions*

The sharp 7.6-fold jump in NAK_rate between Hour 5 and Hour 6 aligns with the diurnal traffic rhythms reported in Ref. [1]. This transition period—corresponding to the onset of business hours in Turkey (UTC+3)—marks the convergence of overnight batch processes completing and interactive daytime traffic spinning up. The magnitude of variation substantially exceeds the 2–3× swings typically seen in laboratory studies [6], underscoring that production environments exhibit more pronounced temporal dynamics than controlled testbeds can reproduce. The secondary peak at Hour 10 coincides with mid-morning workflow intensification, a pattern consistent with enterprise network utilization literature.

Tuesday's emergence as the worst-performing day (11.5× higher NAK_rate than Monday) likely reflects the accumulation of deferred Monday tasks combined with full operational capacity. The Tuesday-morning hotspot (Hours 6–10) visible in the heatmap analysis suggests that the interaction between daily and hourly cycles is multiplicative rather than additive. These findings extend the seasonal decomposition work on video streaming in Ref. [18], demonstrating that SRT transport metrics exhibit analogous cyclical structures amenable to time series forecasting.

The weekend advantage (5.1× lower NAK_rate) carries direct implications for scheduling. Unlike laboratory setups that maintain constant load [5], production deployments experience the full range of business-cycle-driven traffic swings. The clear 7-day periodicity confirmed through ACF analysis (significant at lags 7 and 14) lays the groundwork for SARIMA-based forecasting, extending the ARIMA approaches applied to general streaming traffic in Ref. [4]. The PACF—AR (1) plus seasonal AR (7)—indicates that both day-to-day persistence and weekly cyclical effects should be incorporated into predictive models.

The strong correlation between CPU utilization and NAK_rate ($r$ = 0.617, 95% CI [0.583, 0.648]) identifies CPU as the strongest system-level correlate of SRT transport variation. This finding is broadly consistent with the resource–quality relationships reported for WebRTC in Ref. [7], though the specific magnitude differs due to protocol-level processing differences. From a practical standpoint, the association suggests that CPU monitoring can serve as an early-warning indicator for periods of elevated packet loss risk: threshold alerts on CPU could prompt preemptive investigation of streaming quality before user-visible degradation occurs. The CPU–network receive correlation ($r$ = 0.677) confirms that incoming traffic volume drives computational load, pointing to traffic-aware scaling as a viable proactive strategy.

Several confounding factors may contribute to the observed CPU–NAK_rate correlation. First, shared temporal patterns—both metrics peak during high-usage hours—may inflate the aggregated correlation through ecological confounding. Second, concurrent system processes consuming CPU independently of SRT packet processing could produce spurious associations. Third, network-level congestion may simultaneously increase both CPU interrupt handling and packet retransmission rates. Establishing causal directionality—whether CPU saturation degrades SRT performance, or whether SRT retransmission load elevates CPU—would require controlled experiments varying CPU load independently, Granger causality analysis, or instrumented kernel-level measurements. We interpret the correlation as indicating a monitoring relationship suitable for operational alerting rather than a confirmed mechanistic causal pathway.

Perhaps the most notable finding about SRT metric relationships is the independence of NAK_rate from latency-related metrics ($r \approx 0.00$ with RTT_avg, RTT_std, and buffer_variance at the raw 30-s resolution). There is a plausible mechanistic explanation for this. SRT's ARQ mechanism triggers NAK packets based on sequence-number gaps rather than round-trip time values: when the receiver detects a missing packet in the sequence, it issues a retransmission request regardless of the current RTT. Consequently, instantaneous retransmission demand (NAK_rate) reflects loss events, while RTT reflects propagation and queuing characteristics—two largely independent physical processes at the packet timescale. The emergence of correlation at longer aggregation intervals ($r$ = 0.229 at 15-min resolution) suggests that both metrics share common network-load drivers without implying a direct causal link at the packet level.

This dichotomy has practical implications for adaptive streaming design. Optimization strategies targeting latency reduction may have minimal impact on packet loss, and vice versa. Protocol developers should consider implementing separate control loops for these independent degradation pathways, consistent with the multi-objective optimization approaches explored in Rao *et al.* [11].

The moderate RAM–RTT_avg correlation ($r$ = 0.509) suggests that memory pressure is associated with latency elevation, potentially through buffer management overhead or garbage collection pauses. This supports the resource modeling approaches advocated for digital twin frameworks, where accurate simulation requires capturing resource–transport interdependencies [24]. Under normal conditions, Random Access Memory (RAM) utilization was remarkably stable (SD = 1.02%), indicating that memory is not the binding constraint for SRT performance—though the 66.30% peak during the December 29 anomaly shows that extreme traffic surges can push memory utilization well beyond its typical range.

The controlled attack experiments revealed that SRT infrastructure is vulnerable at surprisingly low attack intensities. The observation that 500 pps—roughly matching the legitimate stream rate (0.97×)—caused CPU increases of 176%–806% places the vulnerability threshold well below typical DDoS volumes, which commonly exceed thousands or millions of packets per second [8]. The computational overhead of SRT's encryption and ARQ mechanisms, which must process every incoming packet regardless of legitimacy, accounts for this sensitivity. The non-linear jump to 1348% CPU elevation at ~1000 pps hints at the onset of processing queue saturation.

Equally important is what the attacks did not do. SRT transport metrics barely moved during any of the 6 sessions: NAK_rate stayed within −13% to +4% of baseline, and the dropped packet counter shifted by only −39 to +13 packets. The Automatic Repeat Request mechanism effectively masked the system-level distress at the transport layer, meaning that operators relying solely on SRT statistics would have no indication that an attack was underway. This finding makes a strong case for multi-metric monitoring that combines CPU and transport indicators.

The signature framework—CPU-only events for locally-generated attacks versus CPU-plus-network events for external anomalies—offers a practical starting point for automated classification. The December 29 anomaly, with CPU (97.7%) and network (44.66 Mbps) spikes co-occurring within the same two-minute window, produced a qualitatively different pattern from the 6 controlled attacks and exceeded the worst controlled test by a factor of 3.7 in CPU impact.

A caveat is in order regarding the detection thresholds reported here. The *Z*-score criterion (CPU > 3.16%, corresponding to $3\sigma$ above the mean) is specific to the monitored server's hardware configuration and workload profile. Deployments with different CPUs, encoding settings, or concurrent workloads will exhibit different baseline distributions and therefore different optimal thresholds. We recommend establishing site-specific baselines during a minimum 7-day characterization period before deploying automated detection, with periodic recalibration to account for configuration changes or workload evolution.

For streaming operators, the temporal patterns identified here translate into concrete scheduling guidance. Resources should be proactively scaled during Tuesday mornings (06:00–10:00 local time), when degradation risk peaks. Maintenance windows are best placed during weekend periods—Saturday afternoons in particular—when network conditions are most favorable. CPU utilization sustained above the *Z*-score threshold (approximately $3\sigma$ above the site-specific mean) should trigger alerting, as this level reliably signals impending quality degradation. Real-time dashboards should present CPU alongside NAK_rate for comprehensive quality visibility.

For security teams, the study argues for multi-metric correlation analysis over single-metric thresholds. The signature framework—CPU-only for internal/local threats, CPU-plus-network for external threats—enables rapid event classification and appropriate escalation. Rate limiting at 100–200 pps per source IP provides protection against the demonstrated low-threshold vulnerability while minimizing disruption to legitimate traffic. Monitoring should be intensified during peak-degradation windows (Tuesday mornings), when attack impacts may be masked by or misattributed to normal traffic stress.

For protocol developers, the independence of NAK_rate from latency metrics points to the need for separate control mechanisms for packet loss and latency optimization. Time-aware adaptation that anticipates identified temporal patterns could preemptively adjust encoding parameters during high-risk hours. Buffer management algorithms should account for the afternoon latency peaks (Hours 14–17) through increased buffer allocation. The tight RTT–jitter correlation ($r$ = 0.891) indicates that smoothing algorithms targeting jitter reduction will simultaneously improve latency consistency.

This study extends the SRT literature along several axes. The 96-day monitoring duration exceeds, to our knowledge, the longest prior SRT study (approximately one week [6]) by more than an order of magnitude, enabling detection of weekly cyclical patterns and long-term trends that shorter campaigns cannot capture. The sample volume of 262,672 SRT measurements represents a roughly 100-fold increase over typical laboratory studies, providing the statistical power needed to detect subtle temporal effects.

Methodologically, the first application of time series decomposition to SRT transport metrics extends the forecasting approaches applied to related domains. While RNN-based forecasting was applied to Voice over Internet Protocol (VoIP) quality and ARIMA was used for general streaming traffic, no prior work had characterized the temporal structure of SRT-specific variables [3, 4]. The identified weekly seasonality and AR (1) + seasonal AR (7) lag structure provide a concrete starting point for predictive models. The security assessment fills a gap flagged in comprehensive DDoS surveys regarding streaming-protocol-specific vulnerability [8]. And the production deployment distinguishes this work from laboratory testbed studies, lending ecological validity that artificial workloads cannot replicate [5]. The observation that production environments exhibit more pronounced temporal variations (7.6 × hourly) than laboratory reports (2–3×) underscores the importance of field validation for streaming protocol research.

Several limitations warrant acknowledgment. The study was conducted at a single production site in Turkey; regional network characteristics, infrastructure configurations, and traffic patterns may differ elsewhere. Multi-site validation is necessary to assess how well the temporal patterns and correlation structures generalize. The 96-day window spans autumn and winter (October 2025 through January 2026); summer months may yield different demand and utilization patterns not captured here.

The controlled attacks were executed in an isolated laboratory segment physically disconnected from production systems. While this isolation was essential for ethical and operational reasons, real-world attacks may differ in important ways—distributed source IPs, variable intensity over time, and sophisticated evasion techniques not represented in the controlled tests.

The correlational analyses establish statistical associations, not causal relationships. Although the strong CPU–NAK_rate association ($r$ = 0.617) is suggestive, controlled experiments independently manipulating CPU utilization under constant network conditions would be needed to confirm the direction of causation. The findings apply to the specific SRT implementation (libSRT v1.5.4 on a resource-constrained laptop); other implementations or hardware platforms may behave differently.

The December 29 anomaly, while yielding valuable observational data, cannot be attributed to a specific cause without packet capture logs, system audit trails, or ISP-level traffic records. The characterization of this event as exhibiting a distinct signature from the controlled attacks should be read as an observational classification rather than a confirmed attribution. Future monitoring configurations should include continuous lightweight packet capture with retroactive deep inspection triggered by anomaly detection.

The attack experiments were limited to 6 sessions across 2 vectors (SRT-specific and generic UDP floods). A more thorough security assessment would require additional attack types (SYN flood, application-layer attacks), higher intensities, repeated trials for statistical power ($n$ > 3 per type), and varied SRT configurations (different latency settings, encryption modes). The small sample size limits the precision of the mean ± SD estimates and precludes significance in between-type comparisons.

Several promising directions emerge. Multi-site validation across geographically distributed servers would clarify which temporal patterns are region-specific and which are universal. Machine learning forecasting—particularly LSTM and Transformer architectures—could exploit the identified temporal structures for real-time NAK_rate and RTT prediction, enabling proactive rather than reactive quality management. Operationalizing the signature-based classification framework (CPU-only versus CPU-plus-network) into automated anomaly detection tools would translate these findings into production-ready security systems.

On the security side, investigation of SRT-specific countermeasures is warranted: connection validation mechanisms, computational puzzles to raise attacker cost, and protocol-aware rate limiting that distinguishes legitimate SRT control traffic from flood attacks.

## V. Conclusion

This study presents what is, to the best of our knowledge, the most comprehensive temporal and security analysis of SRT protocol behavior reported in the literature. Over 96 days of continuous production monitoring, we collected 262,672 SRT transport metric samples and 5,033,104 system resource measurements, enabling analysis across multiple temporal scales. All four research hypotheses were confirmed through rigorous statistical testing.

The study contributes to the streaming protocol literature in four ways. First, it provides, to our knowledge, the longest temporal characterization of SRT transport metrics—exceeding the longest previously reported study (approximately one week) by more than an order of magnitude—and reveals weekly cyclical patterns amenable to SARIMA forecasting. Second, it offers the first systematic examination of system–SRT correlations, identifying CPU utilization as the strongest correlate of transport variation ($r$ = 0.617 with NAK_rate), while acknowledging confounding factors that preclude definitive causal interpretation. Third, it introduces the first controlled security evaluation of SRT infrastructure, demonstrating vulnerability at 500 pps (roughly matching the legitimate stream rate) with CPU increases of 176–806%. Fourth, it grounds these findings in a production deployment, showing that real-world environments exhibit more pronounced dynamics than laboratory testbeds capture.

Among the principal findings: NAK_rate varied 7.6-fold across hours (Hour 5 min to Hour 6 peak) and 11.5-fold across days (Monday to Tuesday), though with small effect sizes ($\eta^2$ < 0.04) indicating that temporal period explains less than 4% of total variance. The practical significance lies in specific peak windows where NAK_rate approaches the 0.1% professional broadcast threshold. The independence of NAK_rate from latency metrics ($r \approx 0.00$ at raw resolution) reveals that packet loss and latency are orthogonal quality dimensions requiring separate optimization strategies. Multi-scale correlation analysis shows this independence breaks down at coarser aggregation intervals—an ecological correlation effect that must be accounted for in monitoring system design.

A comparative anomaly detection analysis demonstrated that adaptive $Z$-score methods ($3\sigma$ threshold) detect all identified security events, including the low-rate Attack 3 that evaded the fixed 10% CPU threshold. We recommend statistical anomaly detection over fixed thresholds for production SRT monitoring. Analysis of SRT transport metrics during controlled attacks revealed protocol resilience through the ARQ mechanism, with NAK_rate variation limited to ±13% despite CPU increases exceeding 1000%—a finding that underscores the necessity of multi-metric monitoring combining system resources with transport indicators.

These findings translate into actionable guidance: proactive resource scaling during Tuesday mornings,

weekend maintenance scheduling, site-specific *Z*-score-based CPU alerting, multi-metric correlation for attack detection, and rate limiting at 100–200 pps per source IP. As live streaming continues to expand across broadcast, enterprise, and consumer applications, the temporal patterns, resource correlations, and security vulnerabilities documented here provide a foundation for predictive quality management, automated anomaly detection, and protocol-specific defense mechanisms.

## CONFLICT OF INTEREST

The authors declare no conflict of interest.

## AUTHOR CONTRIBUTIONS

Ahmed F. K. Koysha designed the study, developed the monitoring infrastructure, collected the data, conducted the statistical analysis, performed the security experiments, and wrote the manuscript; Ferdi Sönmez supervised the research, provided methodological guidance, and reviewed the manuscript; both authors approved the final version.